\documentclass[nofootinbib,prd,twocolumn,showpacs,showkeys,preprintnumbers]{revtex4-1}
\usepackage{hyperref,amssymb,amsmath,mathrsfs,bm,graphicx}
\usepackage[dvipsnames]{xcolor}
\usepackage{comment}
\usepackage{caption}
\usepackage{subcaption}

\begin{document}

\title {Double relativistic master polytrope for anisotropic matter}

\author{P. Le\'on}
\email{pablo.leon@ua.cl}
\affiliation{Departamento de F\'isica, Universidad de Antofagasta, Aptdo 02800, Chile.}

\author{E. Fuenmayor}
\email{ernesto.fuenmayor@ciens.ucv.ve}
\affiliation{Centro de F\'isica Te\'orica y Computacional,\\ Escuela de F\'isica, Facultad de Ciencias, Universidad Central de Venezuela, Caracas 1050, Venezuela}

\author{E. Contreras}
\email{econtreras@usfq.edu.ec}
\affiliation{Departamento de F\'isica, Colegio de Ciencias e Ingenier\'ia, Universidad San Francisco de Quito USFQ,  Quito 170901, Ecuador\\}

\begin{abstract}
We present a detailed analysis of a general relativistic static spherical symmetric distribution in which both the radial and tangential pressures follow a master polytropic equation of state that generalizes the standard treatment and avoids the appearance of singularities in the system.
In particular, we find the corresponding Lane-Emden equation and integrate it for a wide range of values of the parameters involved. We explore
the parameter space with the aim to find the set of parameters leading to reasonable physical solutions. Also, we considered the effect of spherically symmetric perturbations of the matter variables in order to analyze the possible apparition of cracking within the compact distribution. 
\end{abstract}


\maketitle

\section{Introduction}
In the study of self-gravitating matter distributions, either in the Newtonian or relativistic regimes, it is a known fact that we need to provide, by hand, some information about the system under study. This can be done by providing an equation of state that could be related to some local physical properties of interest. In this regard, one particular case that allows dealing with a large variety of physical scenarios is the polytropic equation of state (see Refs. \cite{cha, sch,  2, 3, 7', 9, 8,10, 4a, 4b, 4c, 5, 6, 7, 11, 12, 13, pol1n}), that have been extensively used to study the
stellar structure, for both, Newtonian and general relativistic objects. Indeed, the application of this equation goes from the description of the internal regions of
compact objects  (as white dwarf or neutron stars)  \cite{cha,2, 3} to the dynamics of galaxies \cite{gw}. The outstanding success of the polytropic equation of state lies in the simplicity of the ensuing main equation (Lane-Emden).

For a relativistic matter distribution with isotropic pressure, the polytropic equation of state can be written in two different ways:
\begin{eqnarray}
P &=& K \rho^{1+1/n}, \label{p1} \\
P &=& K \rho_0^{1+1/n},
\end{eqnarray}
where $\rho, \rho_0, n,K$ are the energy density, the baryonic mass density, the polytropic exponent, and the polytropic constant, respectively. It is important to mention that, in the Newtonian limit, there is no difference between $\rho$ and $\rho_0$; therefore,  we have only one polytropic equation of state.

Now, although the isotropy in pressure (Pascalian fluid) is a common assumption in the literature for the study of compact objects, in the last decades there has been a growing and justified interest in the study of matter distributions with local anisotropy in pressures \cite{Bowers, CHH,14, LHTolman1, hmo02, 04, hod08, 8p, hsw08, p1, p2, anis1, anis2, anis3, anis4bis,o1,o4,16, 17, 18, 23, 24, 25, 26, 19,vandalen, Anderson, drago, sad, alford, Dong,CoP,HSS}). 
Under the assumption of local anisotropy of pressures (in the case of spherical symmetry), we can identify two different principal stresses, $P_r$ and $P_{\perp}$. The general formalism to study anisotropic polytropes (Newtonian or relativistic) was developed in \cite{1p,2p,3p} (see also \cite{z2, z3, z4, 5p, 6p, 7p, 8lp, Nasheeha} for more recent studies). In this case, the polytropic equation of state is given by
\begin{eqnarray}
P_r &=& K \rho^{1+1/n_r}, \\
P_{r} &=& K \rho_0^{1+1/n_r}.
\end{eqnarray}
However, since this assumption introduces a new degree of freedom ($P_\perp$), the polytropic equation of state is not enough to solve Einstein’s equations. Thus, it is necessary to provide additional information about the system under consideration \cite{hod08,8p} related to the local physics or restrictions on the metric variables such as the conformally flat \cite{3p, Herrera2001}, the Karmarkar condition  \cite{karmarkar} or the complexity factor \cite{VC} which are summarized by the statement of a particular form that takes the anisotropic function $\Delta \equiv P_\perp - P_r$ \cite{maurya, tello2, tello3, Nunez, tello5}. Some years ago, it was developed a general formalism to incorporate anisotropy in polytropic Newtonian and relativistic spheres \cite{1p,2p}, using a heuristic strategy that assumes an anisotropy function proportional to the gravitational force present in the hydrostatic equilibrium equation \cite{Cosenza}. Very recently, a novel strategy was adopted to solve the same problem, based on the plausible assumption that both principal stresses satisfy a polytropic equation of state \cite{5p,6p}.

However, in a recent work, \cite{8lp}, the authors showed that
implementing the polytropic equation of state leads to a singular tangential sound velocity at the surface of the distribution for $n>1$. One way to address this problem is to consider a modification of the polytropic equation of state, known as the ``master'' polytropic equation of state \cite{8lp, Nasheeha}, which is given by 
\begin{eqnarray}\label{p-maestro}
P_r &=& K_r \rho^{1+1/n_r}+ \tilde{\alpha}_r \rho - \tilde{\beta}_r, 
\end{eqnarray}
where $\tilde{\alpha}_r$ and $\tilde{\beta}_r$ are constants, allowing us to model material configurations where the density is different from zero (discontinuous) at the boundary. It is worth mentioning that Eq. (\ref{p-maestro}) has been used to describe several cosmological scenarios as the construction of early universe models in addition to being able to be used for various particular cases. A special case occurs when $n=1$, which corresponds to $P_r = K_r \rho^{2}+ \tilde{\alpha}_r \rho - \tilde{\beta}_r$ that has been used to describe possible compact anisotropic charged objects \cite{45}. Another interesting case results when $\beta = 0$ that has been used to find mathematical models of compact objects incorporating the radiation factor \cite{46}.
Note that, although (\ref{p-maestro})  is not an equation of state that comes from some known thermodynamic process, there are a couple of reasons to consider it a suitable approach in the context of the study of self-gravitating spheres. As claimed in Ref. \cite{8lp}, the master equation avoids the appearance of a singularity in the tangential sound velocity when $\alpha=\beta=0$ (the standard polytropic equation ) with $1<\gamma<2$. 
Also, in a very recent article \cite{LN2}, it was explored the physical acceptability conditions for anisotropic compact relativistic matter configurations considering the master polytrope and a heuristic anisotropy \cite{Cosenza}, and certainly, several models emerging from this study could represent real astrophysical compact objects.

In this work, we will analyze the internal structure of a general realistic self-gravitating matter distribution where not only the radial but the tangential pressure satisfies a master polytropic equation of state \cite{6p}, namely 
\begin{eqnarray}\label{pta}
P_\perp &=& K_\perp \rho^{1+1/n_\perp}+ \tilde{\alpha}_\perp \rho - \tilde{\beta}_\perp.
\end{eqnarray}
At this point, a couple of comments are in order. First, note that after demanding (\ref{pta}), the system is closed, and not any extra condition is required. To be more precise, we do not require any ad hoc assumption for the anisotropy function which in some cases seems to be artificial. Second,
although Eq. (\ref{pta}) can be written in terms of the baryonic mass density also, we shall restrict here to the case in which the master polytropic equation of state is expressed in terms of the energy density (since the analysis in both cases is very similar). In addition to this, polytropes based on the total energy density seem to be more viable than those with baryonic density, if also they present small positive local anisotropies \cite{LN2}. 

In order to show here a complete study regarding the response of these systems to perturbations, we shall use the scheme developed in \cite{efl} to study the apparition of cracking in our model. The concept of cracking was initially introduced in \cite{cn} (for more recent works see \cite{cnbis, cn1, cn2, cn4, cn6, mardam, 4p}) to describe a fluid distribution at the exact moment in which the system departures from equilibrium. At this point, radial forces will appear in the matter distribution. We say that there is cracking if the radial force is directed inward in the inner part of the sphere and reverses its sign beyond some value of the radial coordinate. In the opposite case, we say that there is an overturning. The apparition of cracking (overturning) is closely related to local anisotropy in pressures.  In fact, in \cite{cn1}, it was shown that cracking results only in the cases in which the local anisotropy is perturbed and it may lead to drastic changes in the evolution of the system  \cite{cn1}. Besides, it is important to mention that the concept of cracking is related to the problem of structure formation \cite{cn6, cn7}. In this work, it has been assumed that spherical symmetry is preserved by perturbations on the matter variables. It should be mentioned that it is possible to consider systems that, although initially spherically symmetric, are submitted to perturbations deviating the system from this symmetry. It has been studied the fact that perturbations deviating the system from spherical symmetry may induce transverse cracking in the fluid distributions \cite{cn2}.

This manuscript is organized as follows: In the next section, we present the Einstein field equations and conventions for a spherically symmetric distribution with local pressure anisotropy. In section III, we give a brief summary of the general perturbation scheme to study the appearance of cracking for our adopted master polytrope models. Section IV is dedicated to briefly reviewing the main aspects of anisotropic polytropes. We devote section V to formulating the double master polytrope. In Section VI we compare the physical properties of the standard double polytrope with those of the double master polytrope and we study the conditions of physical acceptability. In section VII, we will analyze the appearance of cracking in our model. Finally, in the last section, we will discuss all our results.


\section{The Einstein field equations}

Let us consider a static, spherically symmetric distribution of an anisotropic fluid bounded by a surface $\Sigma$.  In like--Schwarzschild coordinates, the metric is given by
\begin{eqnarray}\label{metric}
ds^{2}=e^{\nu}dt^{2}-e^{\lambda}dr^{2}-r^{2}(
d\theta^{2}+\sin^{2}\theta d\phi^{2}),
\end{eqnarray}
where $\nu$ and $\lambda$ are functions of $r$ that satisfy regularity conditions at $r = r_c = 0$: $\lambda_c = 0$ and $\nu'_c =\lambda'_c=0$ (where the subscript $c$ refers to the center of the distribution).

The matter content of the sphere is described by a non-Pascalian fluid represented by the energy-momentum tensor 
\begin{eqnarray}
T_{\mu\nu}=(\rho+P_{\perp})u_{\mu}u_{\nu}-P_{\perp}g_{\mu\nu}+(P_{r}-P_{\perp})s_{\mu}s_{\nu},
\end{eqnarray}
where, 
\begin{eqnarray}
u^{\mu}=(e^{-\nu/2},0,0,0),
\end{eqnarray}
is the four-velocity of the fluid, $s^{\mu}$ is defined as
\begin{eqnarray}
s^{\mu}=(0,e^{-\lambda},0,0),
\end{eqnarray}
with the properties $s^{\mu}u_{\mu}=0$, $s^{\mu}s_{\mu}=-1$ (we are assuming geometric units $c=G=1$). The metric (\ref{metric}), has to satisfy the Einstein field equations, which are given by
\begin{eqnarray}
\rho&=&-\frac{1}{8\pi}\bigg[-\frac{1}{r^{2}}+e^{-\lambda}\left(\frac{1}{r^{2}}-\frac{\lambda'}{r}\right) \bigg],\label{ee1}\\
P_{r}&=&-\frac{1}{8\pi}\bigg[\frac{1}{r^{2}}-e^{-\lambda}\left(
\frac{1}{r^{2}}+\frac{\nu'}{r}\right)\bigg],\label{ee2}
\end{eqnarray}
\begin{equation}
P_{\perp}=\frac{1}{8\pi}\bigg[ \frac{e^{-\lambda}}{4}
\left(2\nu'' +\nu'^{2}-\lambda'\nu'+2\frac{\nu'-\lambda'}{r}
\right)\bigg]\label{ee3},
\end{equation}
where primes denote derivative with respect to $r$.

Outside the fluid distribution, the spacetime is given by the Schwarzschild exterior solution, namely
\begin{eqnarray}
ds^{2}&=&\left(1-\frac{2M}{r}\right)dt^{2}-\left(1-\frac{2M}{r}\right)^{-1}dr^{2}\nonumber\\
&&-r^{2}(d\theta^{2}+\sin^{2}\theta d\phi^{2}).
\end{eqnarray}

Furthermore, we require the continuity of the first and  the second fundamental form across the boundary surface $r=r_{\Sigma}=\rm constant$, which implies,
\begin{eqnarray}
e^{\nu_{\Sigma}}&=&1-\frac{2M}{r_{\Sigma}},\label{nursig}\\
e^{-\lambda_{\Sigma}}&=&1-\frac{2M}{r_{\Sigma}}\label{lamrsig}\\
P_{r_{\Sigma}}&=&0 \label{prsig},
\end{eqnarray}
where the subscript $\Sigma$ indicates that the quantity is evaluated at the boundary surface $\Sigma$.

From the radial component of the conservation law,
\begin{eqnarray}\label{Dtmunu}
\nabla_{\mu}T^{\mu\nu}=0,
\end{eqnarray}
one obtains the generalized Tolman--Oppenheimer--Volkoff equation for anisotropic matter which reads,
\begin{eqnarray}\label{TOV}
\mathcal{R} \equiv P_{r}'+\frac{\nu'}{2}(\rho +P_{r})-\frac{2}{r}(P_{\perp}-P_{r})=0,
\end{eqnarray}
where $\mathcal{R}$ defines the total radial force on each fluid element. Alternatively, by using
\begin{eqnarray}
\nu'=2\frac{m+4\pi P_{r}r^{3}}{r(r-2m)},
\end{eqnarray}
where the Misner ``mass'' function $m$ is defined through
\begin{eqnarray}
e^{-\lambda}=1-2m/r\label{mf},
\end{eqnarray}
or, equivalently as
\begin{eqnarray}\label{funcionm}
\frac{dm}{dr}=4\pi r^2 \rho \quad \Rightarrow \quad m=4\pi \int_{0}^{r} \tilde{r}^2 \rho d\tilde{r},
\label{mf2}
\end{eqnarray}
we may rewrite Eq. (\ref{TOV}) in the form
\begin{eqnarray}\label{TOV2}
P_{r}'=-\frac{m+4\pi r^{3} P_{r}}{r(r-2m)}(\rho+P_{r})+\frac{2}{r}\Delta, \label{TOVb}
\end{eqnarray}
where
\begin{eqnarray}\label{delta}
\Delta=P_{\perp}-P_{r},
\end{eqnarray}
measures the anisotropy of the system. The term $\frac{2}{r}\Delta$ is known as the anisotropic strength, which competes to shape the reacting pressure gradient, and the first term on the right of (\ref{TOV2}) clearly represents the gravitational force \cite{Bowers}. It is important to note that $\mathcal{R}$ has dimensions of force per unit volume so, it is the total force per unit volume over each fluid element. Now, if the system is in equilibrium, these contributions cancel out so that $\mathcal{R}=0$ (a vanishing total force). Nevertheless, in the case of generating (via perturbations) a dynamic instability, we will obtain a nonzero local contribution representing the hydrodynamic force on each fluid element. 

We emphasize that it is equivalent to solve the Einstein system (\ref{ee1})–(\ref{ee3}) or to integrate the structure equations (\ref{funcionm})–(\ref{TOV2}). In the first case, we obtain the physical variables $\rho(r)$, $P_{r}(r)$ and $P_{\perp}(r)$ given the metric functions $\lambda (r)$ and $\nu(r)$, while in the second approach, we integrate the structure Eqs. (\ref{funcionm})–(\ref{TOV2}) providing two equations of state or other physical conditions. These equations of state that involve the pressure thermodynamic variables are used together with the coupling conditions (\ref{nursig}), (\ref{lamrsig}), and (\ref{prsig}), leading to a system of differential equations for $\rho(r)$ which can be solved to complete the inner structure of a self-gravitating relativistic compact object.

As already mentioned in the previous section, in order to integrate (\ref{TOV2}), we shall need additional conditions. The main objective of this work is to build a model consisting of a master equation of state for both, radial and tangential pressures, i.e., the master double polytrope \cite{6p}. This constitutes our main ansatz in order to solve the system of equations. Before doing so, in the next section, we introduce the perturbative scheme used to study cracking in our models.

\section{Perturbation scheme}\label{section3}

The concept of cracking was introduced to describe the behavior of a fluid distribution just after its departure from equilibrium. Specifically, we consider perturbations that lead to a departure from the equilibrium of the system on a time scale that is smaller than the hydrostatic one, which is the typical time in which a system reacts to a perturbation of its equilibrium.  In this regard, the perturbation scheme we present in this section to analyze the occurrence of cracking consists, basically, of taking a snapshot of the system just after leaving equilibrium. For this reason, we consider that the perturbations are time-independent. It is important to emphasize that cracking represents only the tendency of the system just after leaving hydrostatic equilibrium, what actually happens next depends on a full dynamical treatment of Einstein's equations. In what follows we will summarize the perturbation scheme proposed in \cite{efl}.

Let us start with a spherical anisotropic relativistic fluid distribution satisfying the generalized hydrostatic equilibrium equation (\ref{TOV}). Besides, the pressures are considered as functions of the energy density and the anisotropic function, i.e.  
\begin{equation}
P_r(\rho,\Delta), \quad P_{\perp}(\rho,\Delta).
\end{equation}
Now, to study the appearance of cracking, we shall perform perturbations of the energy density and the local pressure anisotropy 
\begin{eqnarray}
    \tilde{\rho} &=& \rho + \delta \rho, \\
    \tilde{\Delta} &= & \Delta + \delta \Delta,
\end{eqnarray}
where $\delta \rho$ and $\delta \Delta$ indicate small perturbations that may depend on $r$. It is worth emphasizing that, the assumption on the time scale does not guarantee that the system will lose its symmetry or reach another equilibrium stage after the perturbation. In order to figure out if this is the case, it is necessary to solve the time-dependent Einstein's equations for a period of time greater than the hydrostatic time scale, which is clearly out of the scope of this work. Thus, we can write the perturbed quantities (up to first order) like,
\begin{eqnarray}
    P_r \; &\longrightarrow& \; \tilde{P}_r = P_r + \left(\frac{\partial\tilde{P}_r}{\partial \tilde{\rho}}\right)_{\begin{array}{c}
         \tilde{\rho}=\rho  \\
         \tilde{\Delta}= \Delta 
    \end{array}} \delta\rho \nonumber \\ && \hspace{0.5cm}+\left(\frac{\partial\tilde{P}_r}{\partial \tilde{\Delta}}\right)_{\begin{array}{c}
         \tilde{\rho}=\rho  \\
         \tilde{\Delta}= \Delta 
    \end{array}} \delta\Delta,  \\ && \nonumber \\
    m\; &\longrightarrow& \; \tilde{m} = m + \left(\frac{\partial\tilde{m}}{\partial \tilde{\rho}}\right)_{\begin{array}{c}
         \tilde{\rho}=\rho   \\
         \tilde{\Delta}= \Delta 
    \end{array}} \delta\rho, \\
    \Delta \; &\longrightarrow&\; \tilde{\Delta} = \Delta + \delta \Delta.
\end{eqnarray}
Now, let us assume
\begin{eqnarray}\label{funcionP}
    \tilde{P}_r = (1+\delta \phi) P_r \; , \quad |\delta \phi|<<1,
\end{eqnarray}
where $\delta \phi$ is a constant that ensures that the radial pressure maintains the same functional behavior. As a consequence, we have
\begin{eqnarray}
   \frac{d\tilde{P}_r}{dr} = (1+\delta \phi) \ \frac{dP_r}{dr} \quad \Longrightarrow \quad \delta P'_r = P'_r \ \delta \phi. 
\end{eqnarray}
Moreover, this implies a restriction over the perturbation functions, which is 
\begin{eqnarray*}
   && \left(\frac{\partial\tilde{P}_r}{\partial \tilde{\rho}}\right)_{\begin{array}{c}
         \tilde{\rho}=\rho  \\
         \tilde{\Delta}= \Delta 
    \end{array}} \delta \rho + \left(\frac{\partial\tilde{P}_r}{\partial \tilde{\Delta}}\right)_{\begin{array}{c}
         \tilde{\rho}=\rho  \\
         \tilde{\Delta}= \Delta 
    \end{array}} \delta \Delta = P_r \ \delta \phi ,
\end{eqnarray*}
which leads to 
\begin{eqnarray}
   \delta \rho &=& \left\lbrace \left(\frac{\partial\tilde{P}_r}{\partial \tilde{\rho}}\right)^{-1} \left(P_r \delta \phi - \left(\frac{\partial \tilde{P}_r}{\partial \tilde{\Delta}}\right)\delta \Delta\right) \right\rbrace_{\begin{array}{c}
         \tilde{\rho}=\rho \\
         \tilde{\Delta}= \Delta 
    \end{array}}.
\end{eqnarray}
and in this way (\ref{funcionP}) is satisfied. Then, after perturbation,  we can write 
\begin{eqnarray}
   \tilde{\mathcal{R}}(\tilde{\rho},\tilde{\Delta}) = \mathcal{R}(\rho,\Delta)+\delta \mathcal{R} (\rho,\Delta),
\end{eqnarray}
where
\begin{eqnarray}
   \delta \mathcal{R} &=& \left(\frac{\partial \mathcal{R}}{\partial P_r}\right)_{\begin{array}{c}
         \tilde{\rho}=\rho  \\
         \tilde{\Delta}= \Delta 
    \end{array}}\delta P_r + \left(\frac{\partial \mathcal{R}}{\partial \rho}\right)_{\begin{array}{c}
         \tilde{\rho}=\rho  \\
         \tilde{\Delta}= \Delta 
    \end{array}}\delta \rho \nonumber \\ &+& \left(\frac{\partial \mathcal{R}}{\partial m}\right)_{\begin{array}{c}
         \tilde{\rho}=\rho  \\
         \tilde{\Delta}= \Delta 
    \end{array}} \delta m \nonumber + \left(\frac{\partial \mathcal{R}}{\partial \Delta}\right)_{\begin{array}{c}
         \tilde{\rho}=\rho  \\
         \tilde{\Delta}= \Delta 
    \end{array}}\delta \Delta \nonumber \\ &+& \left(\frac{\partial \mathcal{R}}{\partial P'_r}\right)_{\begin{array}{c}
         \tilde{\rho}=\rho  \\
         \tilde{\Delta}= \Delta 
    \end{array}} \delta P'_r.
\end{eqnarray}
Now, in order to avoid a singularity at the center of the distribution we will choose $\delta \Delta = \Delta\, \delta \beta$ with $\delta\ \beta<<1$ a constant. Thus, the total radial force after the perturbations can be written as (see \cite{efl} for more details)  

\begin{eqnarray}\label{Rfinal}
\mathcal{\tilde{R}} &=& \Bigg\{P_r \Bigg[\frac{4\pi r(\rho+ P_r)}{1-2m/r}+(1+G(r))\Bigg(\frac{m+4\pi r^3P_r}{r^2(1-2m/r)}\Bigg)\Bigg] \nonumber \\ &+& \frac{4\pi(\rho+P_r)(1+8\pi r^2 P_r)F_1(r)}{r^2(1-2m/r)^2} +  P'_r\Bigg\}\;\delta \phi \nonumber \\ &-& \Bigg\{ G(r)\Delta(r)\Bigg[\Bigg(\frac{\partial P_r}{\partial \Delta}\Bigg)\Bigg(\frac{m+4\pi r^3P_r}{r^2(1-2m/r)}\Bigg) \Bigg] \nonumber \\ &+& \frac{4\pi(\rho+P_r)(1+8\pi r^2 P_r)F_2}{r^2(1-2m/r)^2} + \frac{2}{r}\Delta(r)\Bigg\} \;\delta \beta .
\end{eqnarray}
where
\begin{eqnarray}
F_1(r) &\equiv& \int^r_0 \Bar{r}^2  G(r) P_r d\Bar{r}, \\
F_2(r) &\equiv& \int^r_0 \Bar{r}^2 G(r) \left(\frac{\partial \tilde{P}_r}{\partial \tilde{\Delta}}\right) \Delta(r) d\Bar{r}, \\
G(r) &\equiv& \left(\frac{\partial\tilde{P}_r}{\partial \tilde{\rho}}\right)^{-1}\; .
\end{eqnarray}

Now, it is clear that the change of sign which has to be present in the total radial force, required for the existence of cracking (or overturning), implies $\mathcal{\Bar{R}} =0$ for some $ r \in (0,r_{\Sigma})$. This leads to
\begin{eqnarray}\label{Gamma}
\delta \phi =  \Gamma \delta \beta, 
\end{eqnarray}
where
\begin{eqnarray}
\Gamma^{-1}  &=& \Bigg\{P_r \Bigg[\frac{4\pi r(\rho+ P_r)}{1-2m/r}+(1+G(r))\Bigg(\frac{m+4\pi r^3P_r}{r^2(1-2m/r)}\Bigg)\Bigg] \nonumber \\ &+& \frac{4\pi(\rho+P_r)(1+8\pi r^2 P_r)F_1(r)}{r^2(1-2m/r)^2}  + P'_r\Bigg\} \Bigg/ \Bigg\{ G(r)\Delta(r) \nonumber \\ &\times&\Bigg[\Bigg(\frac{\partial P_r}{\partial \Delta}\Bigg)\Bigg(\frac{m+4\pi r^3P_r}{r^2(1-2m/r)}\Bigg) \Bigg] \nonumber \\& + &\frac{4\pi(\rho+P_r)(1+8\pi r^2 P_r)F_2}{r^2(1-2m/r)^2} +  \frac{2}{r}\Delta(r)\Bigg\}. \label{crac}
\end{eqnarray}

Note that with equations (\ref{Rfinal})-(\ref{crac}) is possible to evaluate the occurrence of cracking (overturning) in any spherically symmetric system satisfying a barotropic/polytropic equation of state (when a perturbation is introduced using this scheme, the modified (anisotropic) TOV equation does not vanish anymore). Furthermore, if the system satisfies the physical acceptability conditions, it is easy to show that the total radial force will be free of singularities and will be equal to zero at the center of the distribution. Next, we shall expose the basics of the theory of relativistic master polytropes for anisotropic matter.

\section{Relativistic master polytrope for anisotropic matter}

In this section, we shall derive the corresponding relativistic hydrostatic equilibrium equation for a generalized polytropic equation of state, known as the Lane–Emden equation, which essentially constitutes the dimensionless form of Tolman–Oppenheimer–Volkoff expression (\ref{TOV2}) for a polytrope. So, we dedicate this section to discussing the basic set of equations for the relativistic master polytrope for anisotropic matter \cite{8lp, Nasheeha}. The starting point, in this case, is to adopt the following master polytropic equation of state for the radial pressure 
\begin{eqnarray} \label{sq}
P_r = K_r \rho^{1+1/n_r}+\tilde{\alpha}_r \rho -\tilde{\beta}_r. 
\end{eqnarray}
Notice that $K_r$, $\alpha_r$ and $\beta_r$ are non-independent parameters since they are related by the fact that the radial pressure satisfies the matching condition on the surface $\Sigma$ ($P_r = 0$). Then, we have
\begin{eqnarray} \label{rconstantes}
\tilde{\beta}_{r} = K_r \rho^{1+1/n_r}_{\Sigma}+\tilde{\alpha}_r \rho_{\Sigma}.
\end{eqnarray}

Now, defining the variable $\omega$ as
\begin{equation} \label{om}
    \rho = \rho_c \omega^{n_r},
\end{equation}
where $\rho_{c}$ denotes the energy density at the center (from now on the subscript $c$ indicates that the variable is evaluated at the center). For simplicity, we shall define the following constants 
\begin{eqnarray*}
P^0_r = K\rho_{c}^{1+1/n_r}, \quad \tilde{\alpha} = q_0 \alpha, \quad \tilde{\beta}=P^0_{r} \beta, \quad q_0 = P^0_r/\rho_c.
\end{eqnarray*}
Thus, Eq. (\ref{sq}) can be expressed as
\begin{eqnarray}\label{Pr}
P_{r}=P^0_r[\omega^{n_r}(\omega + \alpha_r)-\beta_r].
\end{eqnarray}
Notice that $P^0_{r}$ is not the pressure at the center of the distribution but satisfies
\begin{eqnarray}
P_{rc} = P^0_r (1+\alpha_r - \beta_r). 
\end{eqnarray}
From (\ref{Pr}), we can write 
\begin{eqnarray}
P'_{r}=P^0_r\omega^{n_r}\left(n_r+1 + \frac{\alpha_r\omega{n_r}}{\omega^{n_r}}\right)\omega'.
\end{eqnarray}
Now, by using the matching conditions we can find the following relation
\begin{eqnarray}
\beta_r = \omega^{n_r}_{\Sigma}(\omega_{\Sigma}+\alpha_r),
\end{eqnarray}
which determines the radius of the distribution.

From the above, it can be written the TOV equation (\ref{TOV}) like 
\begin{eqnarray}
&&\left\lbrace 1 + q_0 \left[(\omega + \alpha_r)-\frac{\beta_r}{\omega^{n_r}}\right] \right\rbrace \frac{d\nu}{dr}\nonumber\\ 
&&\qquad\qquad\qquad\qquad\quad = \frac{4\Delta}{\rho_c r \omega^{n_r}} - 2q_0\left[1+n_r+\frac{n_r \alpha_r}{\omega^{n_r}}\right]\omega' ,\nonumber\\
\end{eqnarray}
from where
\begin{eqnarray}
\nu' &=& \left[\frac{4\Delta}{\rho_c r \omega^{n_r}}-2q_0\left[1+n_r+\frac{n_r \alpha_r}{\omega^{n_r}}\right]\omega'\right] \nonumber \\ &\times& \left[ 1+ q_0 \left[(\omega + \alpha_r)-\frac{\beta_r}{\omega^{n_r}}\right]\right]^{-1}.
\end{eqnarray}

Introducing this expression in Eq. (\ref{ee2}) and defining the following dimensionless variables
\begin{eqnarray} \label{av}
r = A x, \quad m = 4\pi \rho_c A^3 \eta, \quad A^2 = \frac{(1+n_r)q_0}{4\pi \rho_c}
\end{eqnarray}
we obtain 

\begin{eqnarray} \label{lemd}
&&\left[\frac{x-2(1+n_r)q_0\eta}{1+q_0(\omega + \alpha_r-\beta_r \omega^{-n_r})}\right]\left[x\frac{d\omega}{dx} \left(1+\frac{n_r \alpha_r}{(1+n_r)\omega}\right) \right. \nonumber \\ &-& \left. \frac{2\Delta}{\rho_cq_0(1+n_r)\omega^{n_r}}\right] + \eta + q_0 x(\omega+\alpha_r)\frac{d\eta}{dx} - q_0\beta_r x^3 =0 \nonumber \\
\end{eqnarray}
and from Eq. (\ref{mf2}) 

\begin{eqnarray}\label{lemd2}
\frac{d\eta}{dx} = x^2 \omega^{n_r}. 
\end{eqnarray}
Equations (\ref{lemd}) and (\ref{lemd2}), form a system of two first-order ordinary differential equations for the three unknown functions $\omega, \eta, \Delta$, depending on the parameters $n_r, \alpha_r, \beta_r, q_0$ attached to the boundary conditions $\eta(0)=0$, $\omega(0)=1$. They correspond to the differential equations that represent the modified Lane-Emden system for the master anisotropic polytrope. Thus it is obvious that in order to proceed further with the modeling of a compact object, we need to provide additional information that depends on the specific physical problem under consideration. The fact that the principal stresses are unequal produces an extra indeterminacy so the introduction of an additional condition to close the system is compulsory \cite{hod08,8p}. For example, in \cite{1p,2p} it was considered a particular ansatz which allowed us to obtain an anisotropic model continually linked with the isotropic case \cite{Cosenza}. Another interesting choice for the local pressure anisotropy was introduced in \cite{5p, 6p} where the main idea was the additional assumption that both principal stresses satisfy polytropic equations of state. Such an approach, based on assuming a ``natural'' description, was called the double relativistic polytrope. Our main objective is the study of the master double polytrope (among other possible interesting master polytropic models). Also, notice that in the limit $\alpha_r,\beta_r \rightarrow 0$, Eqs. (\ref{lemd})-(\ref{lemd2}) constitute the standard system for a matter distribution with a polytropic equation of state, as expected.

It will be useful to calculate the Tolman mass, which is a
measure of the active gravitational mass \cite{LHTolman1}, defined by
\begin{eqnarray}\label{mt1}
m_{T}=\frac{1}{2} r^2 e^{\frac{\nu-\lambda}{2}}\nu '.
\end{eqnarray}
Alternatively, we can calculate the Tolman mass from the equivalent expression \cite{14}, 
\begin{eqnarray}
m_T = e^{\frac{\nu+ \lambda}{2}}(m+4\pi r^3 P_r).
\end{eqnarray}
Now, as before, we define 
\begin{eqnarray}
m_T = 4\pi A^3 \rho_c \eta_T,
\end{eqnarray}
where 
\begin{eqnarray}
\eta_T &=& e^{\nu/2}\sqrt{\frac{x_\Sigma z}{x_\Sigma z -2q_0 (n_r+1)\eta}}\nonumber \\ && \qquad \times (\eta + q_0 x_\Sigma^3 z^3[\omega^{n_r}(\omega+ \alpha_r)-\beta_r]),
\end{eqnarray}
is a dimensionless function and $z =  \frac{x}{x_{\Sigma}}$. On the other hand, using Eq. (\ref{TOV}) we obtain 
\begin{eqnarray}
\frac{d\nu}{dz} = \frac{2}{(\rho+P_r)}\left(\frac{2\Delta}{z}-\frac{dP_r}{dz}\right).
\end{eqnarray}
Then, by integrating this expression we get
\begin{eqnarray}
\nu_{\Sigma}-\nu & = &\int_{z}^{1}\frac{4\Delta}{z[\omega^{n_r}(1+q_0(\omega+\alpha_r))-q_0\beta_r]}dz \nonumber \\ && \qquad-2\int_{\omega}^{\omega_\Sigma} \frac{q_0\omega^{n_r-1}[(n_r+1)\omega + n_r\alpha_r]}{\omega^{n_r}[1+q_0(\omega+\alpha_r)]-q_0\beta_r}d\omega \, ,\nonumber\\ 
\end{eqnarray}
where $\nu_{\Sigma}$ is given by the matching conditions (\ref{nursig}). Now, defining the potential at the surface of the distribution as

\begin{eqnarray}
y = M/r_{\Sigma}
\end{eqnarray}
and the following functions

\begin{eqnarray}
G_1 &=& - \int_{z}^{1}\frac{4\Delta}{z[\omega^{n_r}(1+q_0(\omega+\alpha_r))-q_0\beta_r]}dz, \\
G_2 &=& 2\int_{\omega}^{\omega_\Sigma} \frac{q_0\omega^{n_r-1}[(n_r+1)\omega + n_r\alpha_r]}{\omega^{n_r}[1+q_0(\omega+\alpha_r)]-q_0\beta_r}d\omega, \\
\end{eqnarray}
we can write the dimensionless Tolman mass as

\begin{eqnarray}\label{M-Tolman}
\eta_T &=&  e^{\frac{G_1+G_2}{2}}\sqrt{\frac{(1-2y)x_\Sigma z}{x_\Sigma z -2q_0 (n_r+1)\eta}} \nonumber \\ &\times& (\eta + q_0 x_\Sigma^3 z^3[\omega^{n_r}(\omega+ \alpha_r)-\beta_r]). 
\end{eqnarray}\\

Finally, it is worth noticing that, for the usual polytrope theory, after restoring the speed of light, we can express the stiffness at the center of the matter distribution as \cite{6p}
\begin{eqnarray}
q_{c}\equiv\frac{P_{c}}{\rho_{c}c^{2}},
\end{eqnarray}
implying, in this case, that the Newtonian limit is given by $c \to \infty$, and then we have $q_{c}\to 0$. So consequently, the Newtonian regime for the master polytrope is obtained by performing the limit 
\begin{eqnarray}
q_c = \frac{P_{rc}}{\rho_c} = q_0 (1+\alpha_r-\beta_r) \rightarrow 0,
\end{eqnarray}
which, in general, is equivalent to taking the limit $q_0 \rightarrow 0$. Thus, combining Eqs. (\ref{lemd})-(\ref{lemd2}) and taking the limit $q_0 \rightarrow 0$ reveals that the Newtonian Lane-Emden equation for the master polytrope reads 
\begin{eqnarray}\label{N-MP}
\Bigg[1 &+& \frac{n_r \alpha_r}{(1+n_r)\omega}\Bigg]\frac{d^2\omega}{dx^2} + \left[\frac{2}{x}+\frac{n_r \alpha_r}{1+n_r}\left(\frac{2}{x\omega}-\frac{1}{\omega^2}\frac{d\omega}{dx}\right)\right]\frac{d\omega}{dx} \nonumber \\ &-& \frac{2}{P^0_r(1+n_r)}\left(\frac{1}{x\omega^{n_r}}\right)\left[\frac{d\Delta}{dx}+\frac{\Delta}{x}-\frac{n_r\Delta}{\omega}\frac{d\omega}{dx}\right] = -\omega^{n_r}. \nonumber \\ && 
\end{eqnarray}
Notice that we can recover the usual form of the anisotropic Newtonian Lane-Emden equation by taking the limit $\alpha_r \rightarrow 0$. However, to recover the boundary condition $\omega_{\Sigma}=0$, is necessary to take the limit $\beta_r \rightarrow 0$. It is worth mentioning that Eq. (\ref{N-MP}) can be used to model stellar configurations, in the Newtonian (non-relativistic) regime, that satisfy a master polytropic equation of state. This represents an interesting fact in itself that could be developed in other works.


\section{The double master polytrope}
As we mentioned before, we need to provide further information to integrate the system of equations (\ref{lemd})-(\ref{lemd2}). The neuralgic approach of this work comes from assuming that the tangential pressure ($P_{\perp}$) also satisfies a master polytropic equation of state, so we propose to follow the same strategy as in \cite{5p, 6p} for the master polytrope, this is
\begin{eqnarray}
P_\perp = K_\perp \rho^{1+1/n_\perp} + \tilde{\alpha}_{\perp} \rho - \tilde{\beta}_{\perp},
\end{eqnarray}
which can be written as 
\begin{equation}
 P_\perp = P^0_\perp [\omega^{n_r}(\omega^\theta + \alpha_{\perp})-\beta_{\perp}],
\end{equation}
where $\theta=n_{r}/n_{\perp}$ and as before $ P^0_\perp = K_\perp \rho_c^{1+1/n_\perp}$, $P^0_\perp \alpha_\perp = \tilde{\alpha}_{\perp}\rho_c$ and $P^0_\perp \beta_\perp = \tilde{\beta}_{\perp}$. Thus the anisotropic function $\Delta$ is given by
\begin{eqnarray}
\Delta &=& P^0_\perp [\omega^{n_r}(\omega^\theta + \alpha_{\perp})-\beta_{\perp}] -P^0_{r}[\omega^{n_r}(\omega + \alpha_r)-\beta_r].  \nonumber \\ &&
\end{eqnarray}
Now, since $\Delta(0) = 0 $, it is easy to find
\begin{eqnarray}
P^0_{\perp} = P^0_{r} \left(\frac{1+\alpha_r-\beta_r}{1+\alpha_\perp-\beta_\perp}\right),
\end{eqnarray}
which allow us to write 
\begin{eqnarray} \label{af}
\Delta &=& P^0_r \left[\left(\frac{1+\alpha_r-\beta_r}{1+\alpha_\perp-\beta_\perp}\right)[\omega^{n_r}(\omega^\theta + \alpha_{\perp})-\beta_{\perp}] \right. \nonumber \\ & - &  \omega^{n_r}(\omega + \alpha_r)+\beta_r\Big].
\end{eqnarray}
Introducing this expression in equation (\ref{lemd}) we obtain
\begin{eqnarray} \label{lemd4}
&&\left[\frac{x-2(1+n_r)q_0\eta}{1+q_0(\omega + \alpha_r-\beta_r \omega^{-n_r})}\right]\left[x\frac{d\omega}{dx} \left(1+\frac{n_r \alpha_r}{(1+n_r)\omega}\right) \right. \nonumber \\ &-& \left. \frac{2}{(1+n_r)\omega^{n_r}} \Big\{a(\omega^{n_r}(\omega^\theta + \alpha_{\perp})-\beta_{\perp})  \right. \nonumber \\ & - & \left.  \omega^{n_r}(\omega + \alpha_r)+\beta_r\Big\} \right] \nonumber \\   &+& \eta + q_0 x(\omega+\alpha_r)\frac{d\eta}{dx} - q_0\beta_r x^3 =0,
\end{eqnarray}
where 
\begin{eqnarray}
a \equiv \frac{1+\alpha_r-\beta_r}{1+\alpha_\perp-\beta_\perp}.
\end{eqnarray}\\
Now, we proceed to integrate the system (\ref{lemd4}) and (\ref{lemd2}) numerically by exploring the set of parameters involved with the aim to study the behavior of the matter sector, namely, the density energy, the pressures, the anisotropy, the Tolman mass, and the surface potential.

\begin{figure*}
    \resizebox{0.8 \textwidth}{!}{
    \includegraphics{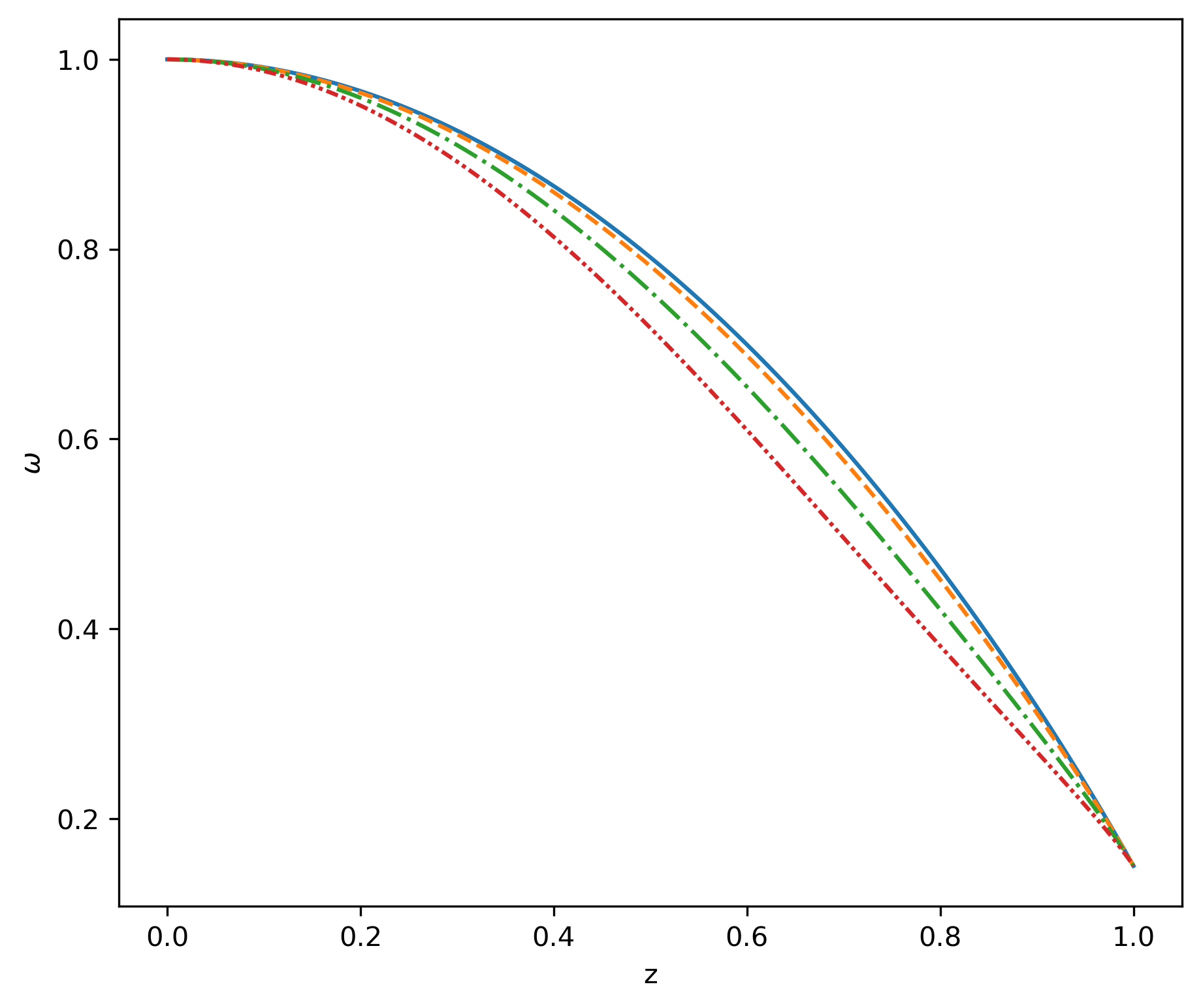} \ \includegraphics{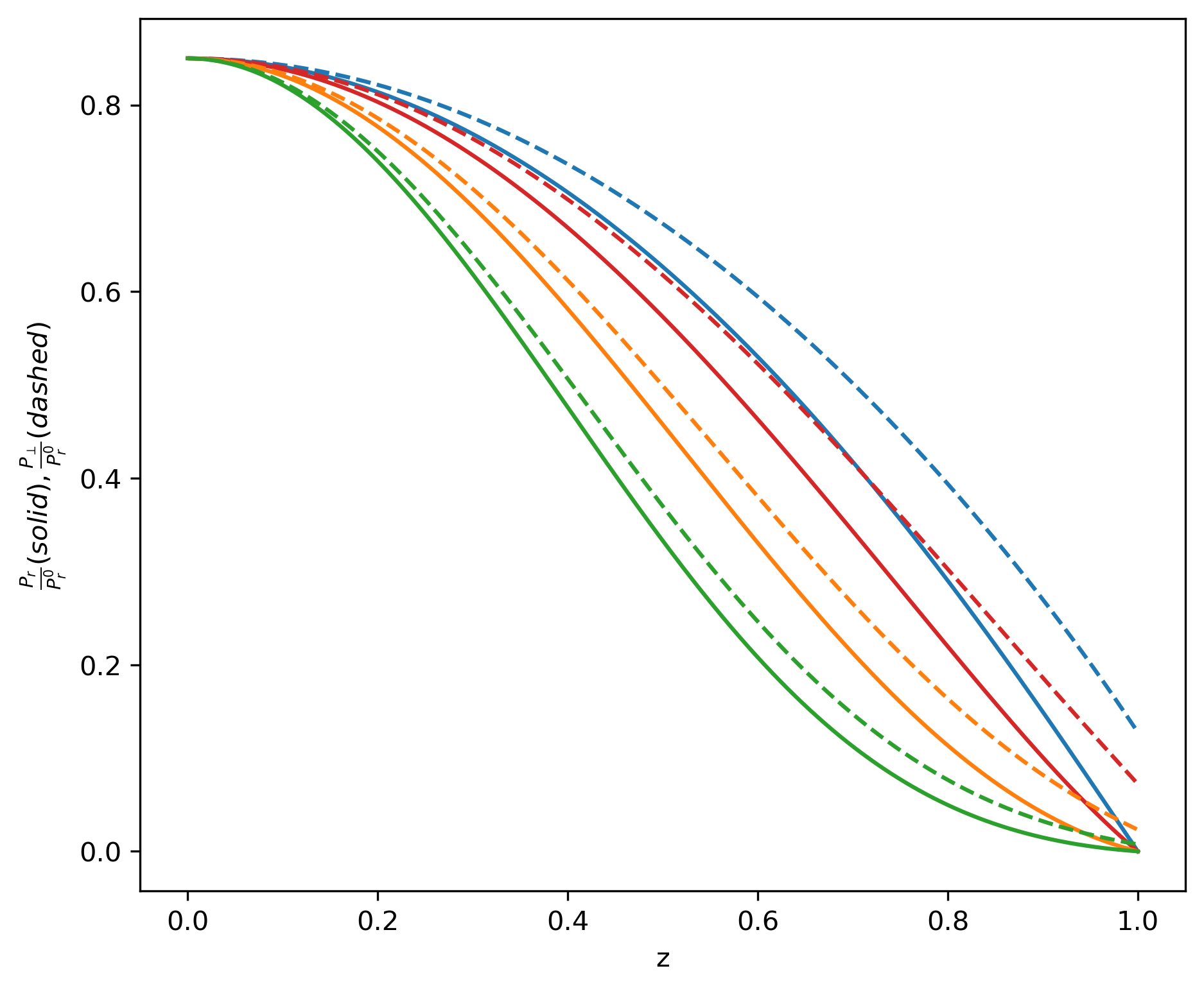}}
    \resizebox{0.8 \textwidth}{!}{
    \includegraphics{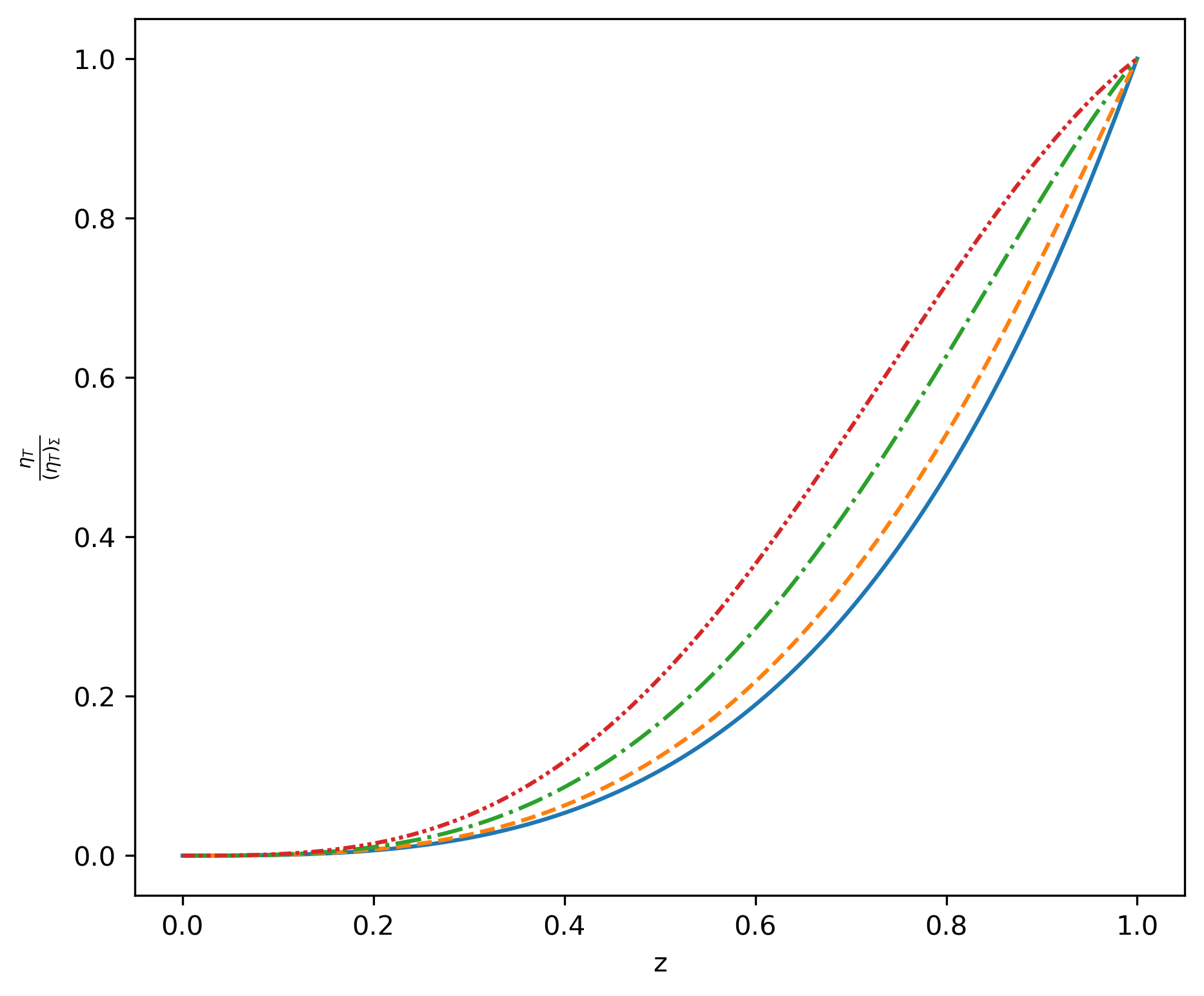} \ 
    \includegraphics{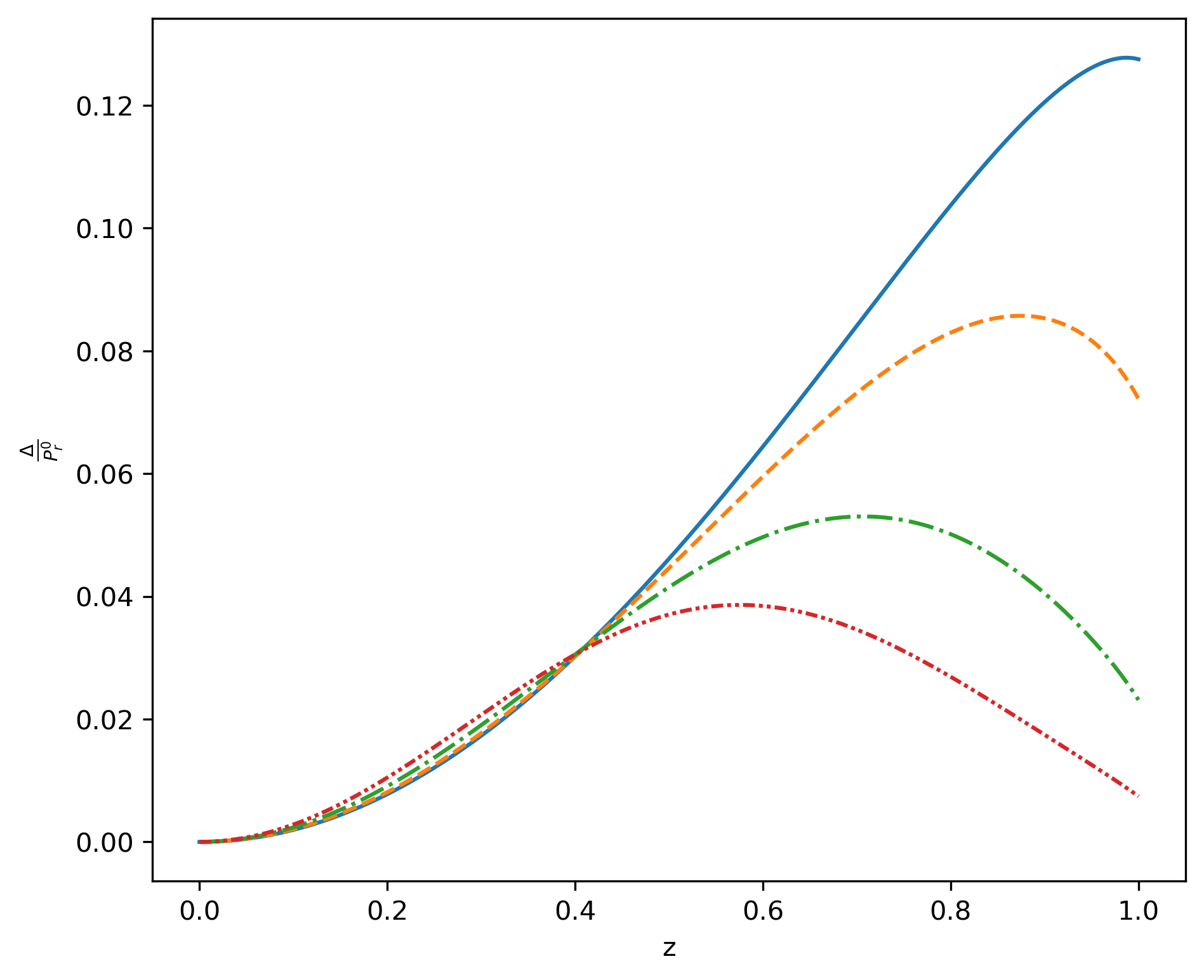}}
    \caption{$w$ (top left), $\eta_T /(\eta_T)_\Sigma$ (bottom left) and $\Delta/P_r^0$ (bottom right) as a function of $z$ for $\theta=0.9$, $q_0 = 0.2$, $\alpha_r =-0.15$, $\beta_r=0$, $\alpha_\perp = 0$, $\beta_\perp=0$ and $n=0.1$ blue (solid) curve, $n=0.4$ orange (dashed) curve, $n=1.0$ blue (dot-dashed) curve, $n=1.4$ red (short double dot-dashed) curve. In the top right its represented $P_r/P^0_r$ (solid) and $P_\perp/P^0_r$ (dashed) as a function of $z$ for $\theta=0.9$, $q_0 = 0.2$, $\alpha_r =-0.15$, $\beta_r=0$, $\alpha_\perp = 0$, $\beta_\perp=0$ and $n=0.1$ blue curve, $n=0.4$ orange curve, $n=1.0$ green curve, $n=1.4$ red curve.}
    \label{fig1}
\end{figure*}

In the first row of Fig. \ref{fig1} we show the behavior of the matter sector, namely the energy density and both, the radial and tangential pressure, as a function of the dimensionless-normalized radial coordinate $z = x/x_\Sigma$ for different choices of the parameters shown in the legend. The matter functions are shown for different values of the index that defines the radial polytrope $n_r$. The junction conditions on the surface (for the radial pressure) determine a single size for the stellar object and depending on the pair ($\alpha_r$, $\beta_r$)  we will have different values for the energy density which is not necessarily continuous on the surface of the object. The distribution of the active gravitational mass towards the inner or outer regions of the sphere depends on the values chosen for these constants, where the blue curve corresponds to the model for which both values are zero. All thermodynamic variables are positive inside the star, reach their maximum at the center and decrease monotonously towards the surface where the radial pressure becomes zero, as expected (while the tangential pressure does not).  Observe that, in our case, the energy density (encoded in $\omega$) does not vanish at the surface of the object. The behavior of the normalized Tolman (``active'' gravitational) mass (\ref{M-Tolman}) and the anisotropy function, useful when analyzing cracking later, (\ref{af}) is shown in the second row (left panel) of Fig. \ref{fig1}, for the parameters chosen in the legend of the figure. The active gravitational mass is zero at the center of the fluid distribution and grows to its maximum value on the surface. In the interior of the fluid distribution, the local anisotropy (Fig. \ref{fig1}, right panel) is an increasing function (as usual) but it modifies its behavior as we approach the surface. This behavior depends radically on the index $n_r$ that determines the structure of the polytrope. We notice that by increasing the index associated with the radial polytrope, the behavior of the anisotropy function changes notably near the surface of the object.

\begin{figure*}
    \resizebox{1.0 \textwidth}{!}{
    \includegraphics{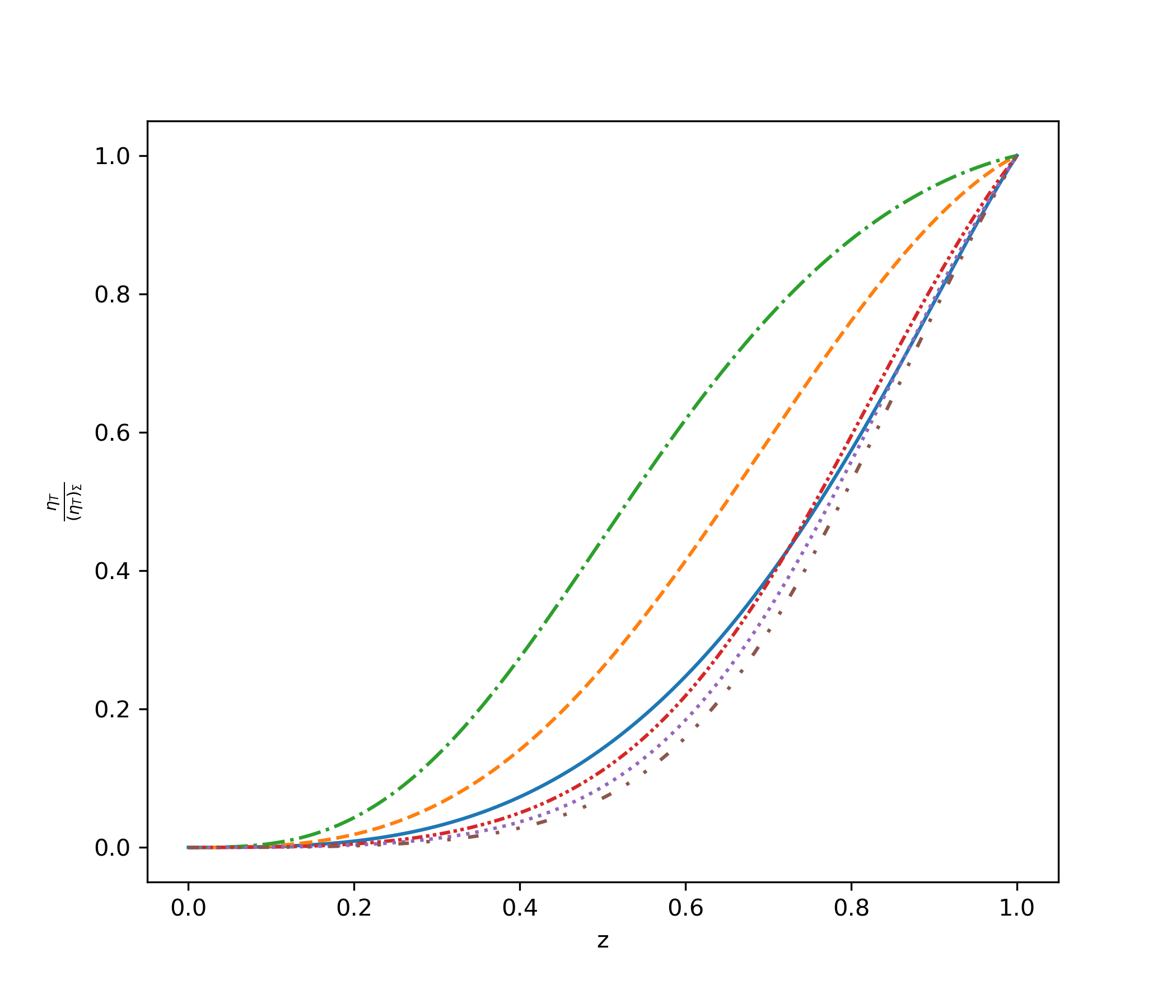} \ \includegraphics{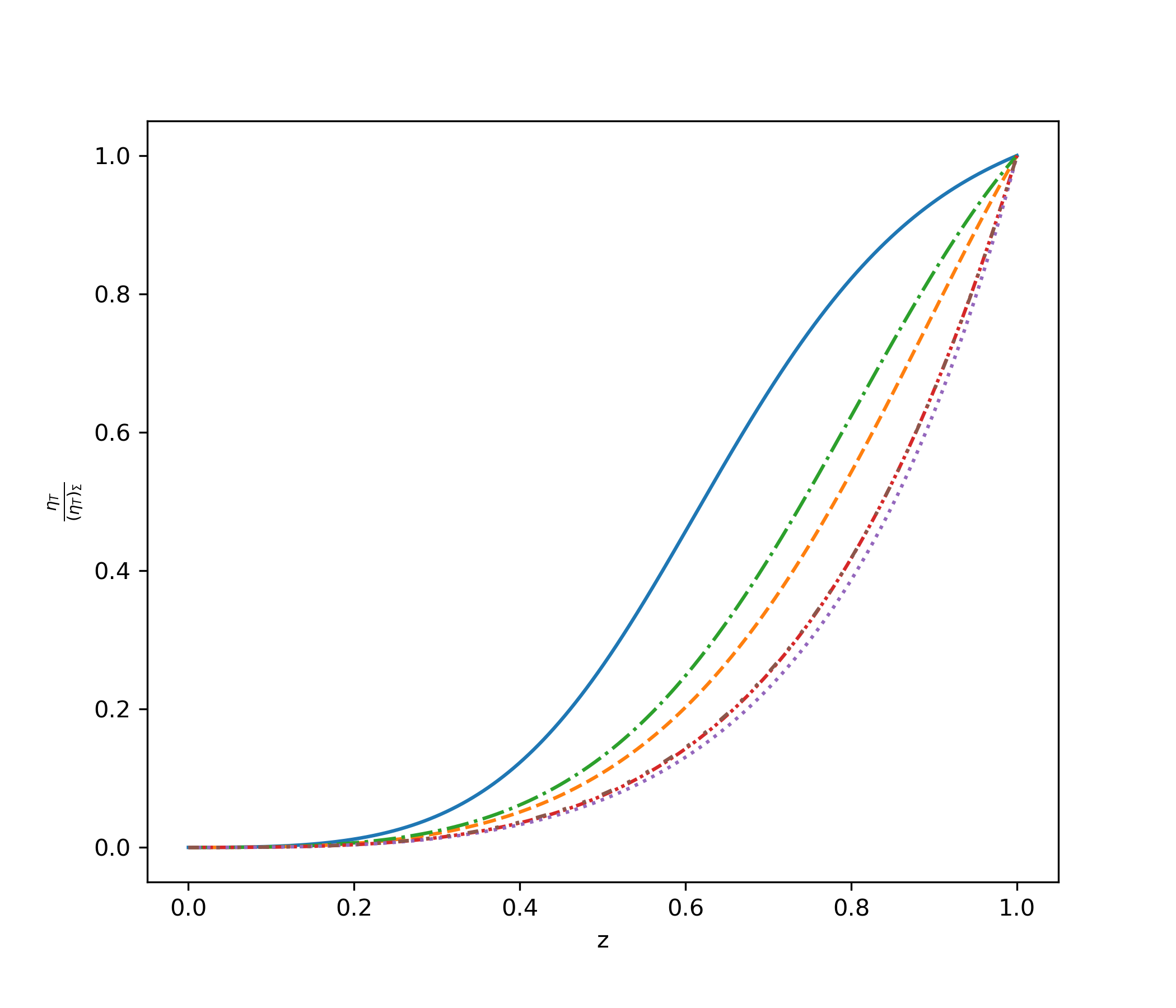}} 
    \caption{Left: $\eta_T /(\eta_T)_\Sigma$ as a function of $z$ for $n=1.5$, $\theta =0.5$, $q_0=0.5$, $\alpha_\perp = 0$, $\beta_\perp=0$ and the pairs $(\alpha_r,\beta_r)=(0,0)$ blue (solid) curve, $(-0.2,0)$ orange (dashed) curve, $(-0.1,0)$ green (dot-dashed) curve, $(0,0.1)$ red (short double dot-dashed) curve, $(0,0.2)$ purple (dotted) curve and $(-0.1,0.1)$ brown (large double dot-dashed) curve. Right: $\eta_T /(\eta_T)_\Sigma$ as a function of $z$ for $\theta=0.5$, $\alpha_r =-0.1$, $\beta_r=0$, $\alpha_\perp = 0.1$, $\beta_\perp=0.001$ and the pairs $(n_r,q_0)=(0.5,0.1)$ blue (solid) curve, $(1.5,0.1)$ orange (dashed) curve, $(2.5,0.1)$ green (dot-dashed) curve, $(1.5,0.6)$ red (short double dot-dashed) curve, $(1.5,0.8)$ purple (dotted) curve and $(1.5,1.0)$ brown (large double dot-dashed) curve.}
    \label{fig2}
\end{figure*}

In the left panel of Fig. \ref{fig2} we represent the normalized gravitational ``active'' mass as a function of $z$ for the values indicated in the figure legend by varying the parameters associated with the radial pressure of the master polytrope, specifically the pair of parameters ($\alpha_r$, $\beta_r$). We note that for the values ($0$, $0$) we return to the usual double polytrope case \cite{6p}. Negative values of $\alpha_r$ produce a shift of the Tolman mass towards the interior regions of the sphere, while increasing positive values of $\beta_r$ produce the opposite effect, shifting the active gravitational mass towards the surface of the compact object. Therefore, it may be inferred from this figure that more stable configurations correspond to more positive values of $\beta_r$ since they are associated with a sharper reduction of the Tolman mass in the inner regions.  In the right panel of the same figure, we show the behavior of the Tolman mass as a function of $z$ but now by changing the pair of parameters ($n_r$, $q_0$). In the case of the master polytrope $q_0$ measures how relativistic the polytrope is (in fact $q_0$ is related to the Newtonian limit). We observe that as $q_0$ increases the mass is concentrated in the outer layers of the sphere effect that could represent the search for stability (avoiding collapse) as the object becomes more compact. As we have already seen $n_r$ represents the type of polytrope we are considering (its radial index). By changing $n_r$ the distribution of the active gravitational mass is different for each master polytrope fluid configuration as usual.

\begin{figure*}
    \resizebox{1.0 \textwidth}{!}{
    \includegraphics{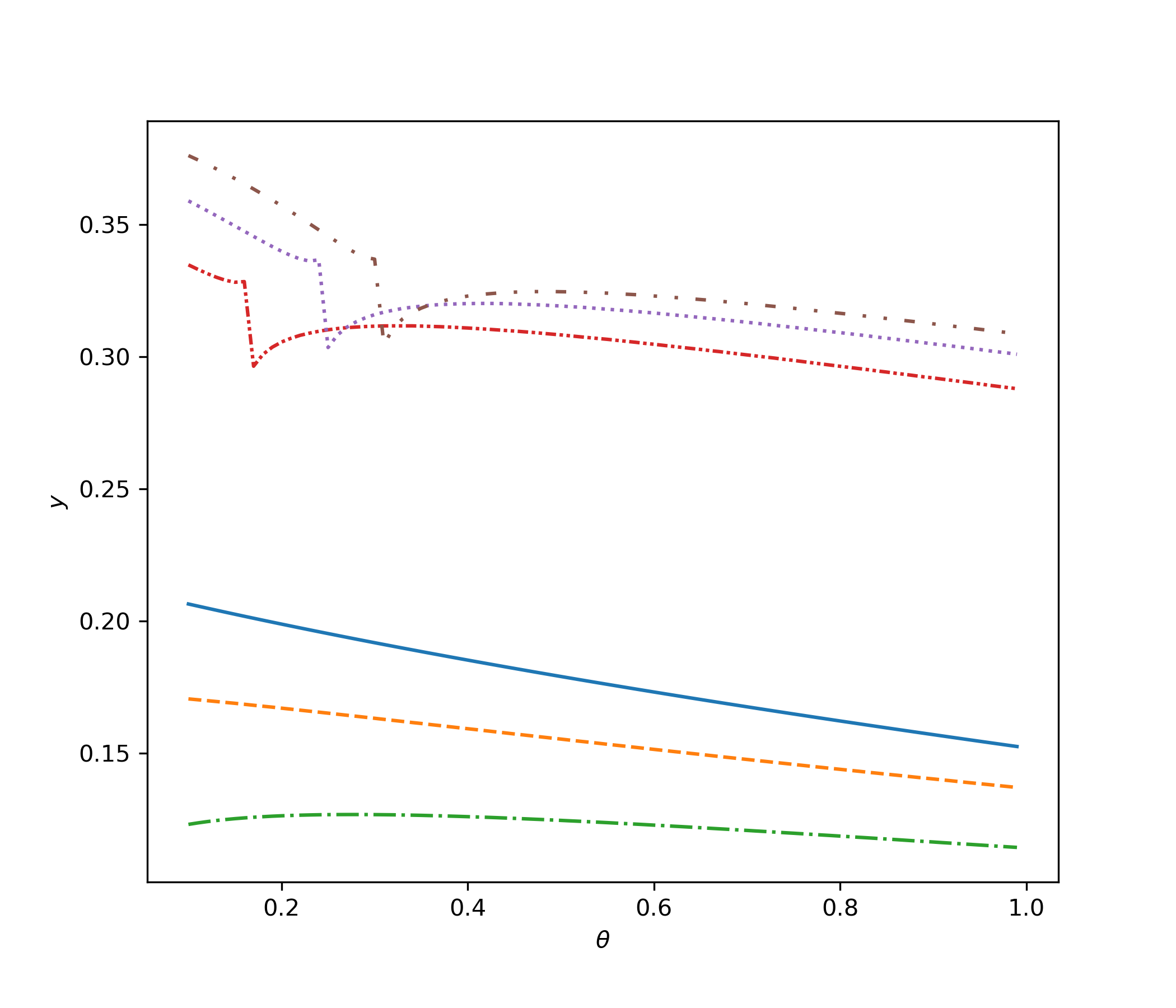} \ 
    \includegraphics{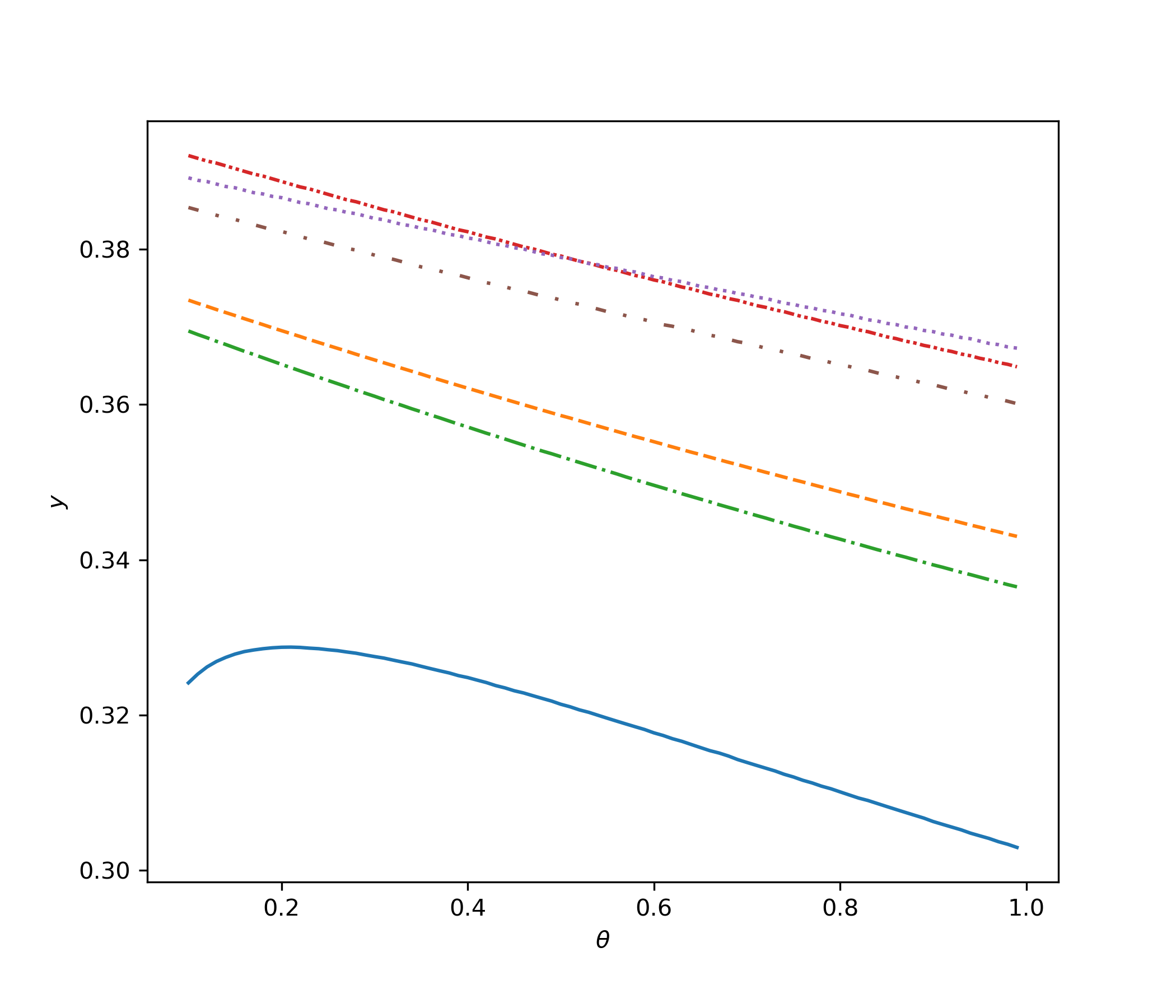}}
    \resizebox{0.47 \textwidth}{!}{
    \includegraphics{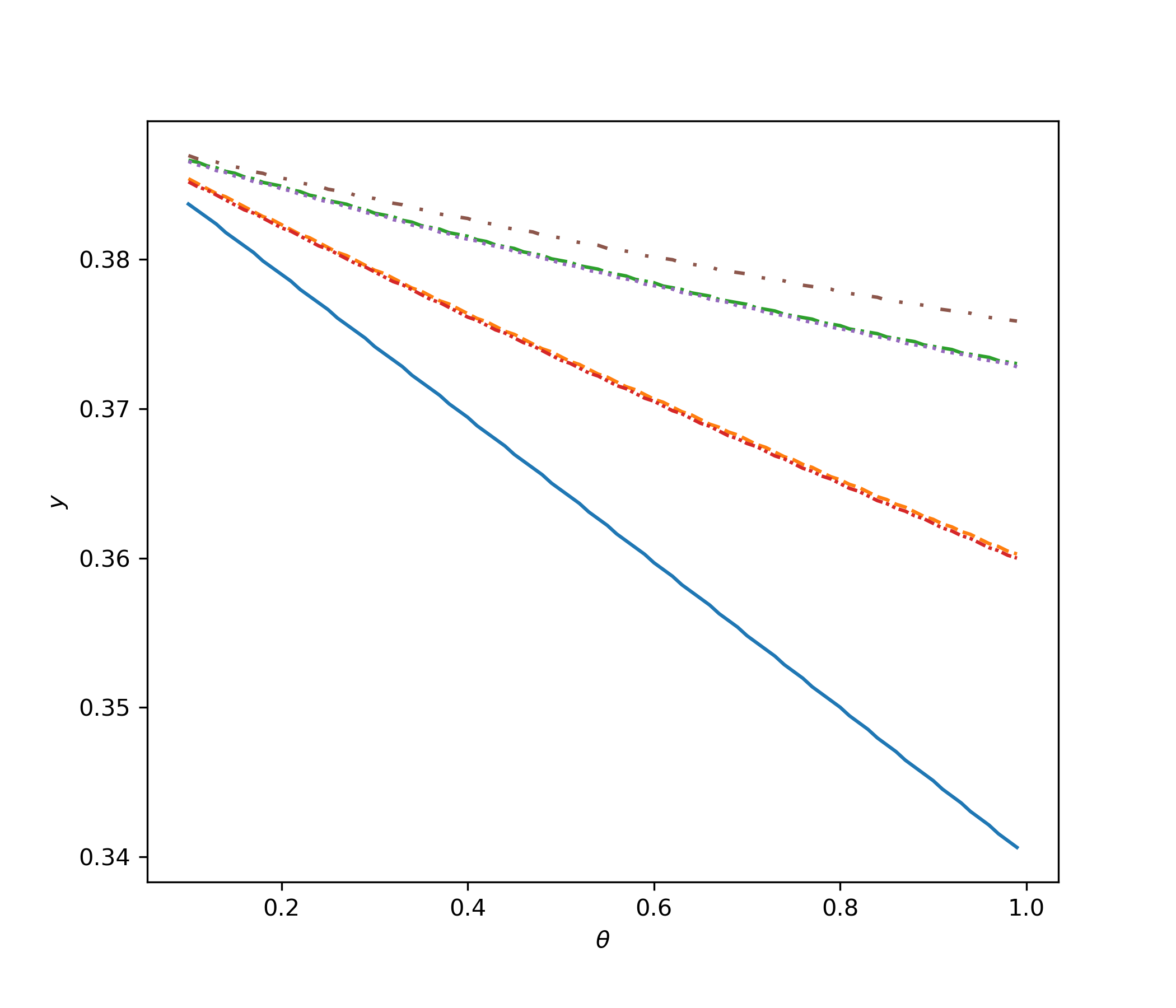} 
    } 
    \caption{Different plots of $y$ vs $\theta$. Top left panel: $\alpha_r = -0.05, \beta_r=0,\alpha_\perp=0.1, \beta_\perp=0.001$ and the pairs $(n_r,q_0)=(0.5,0.1)$ blue (solid) curve, $(1.5,0.1)$ orange (dashed) curve, $(2.5,0.1)$ green (dot-dashed) curve, $(1.5,0.6)$ red ( short double dot-dashed) curve, $(1.5,0.8)$ purple (dotted) curve and $(1.5,1.0)$ brown (large double dot-dashed) curve. Top right panel: $n_r=1.0,q=0.5,\alpha_\perp=0.5,\beta_\perp=0.001$ for $(\alpha_r,\beta_r)=(0,0)$ blue (solid) curve, $(-0.2,0)$ orange (dashed) curve, $(-0.1,0)$ green (dot-dashed) curve, $(0,0.1)$ red ( short double dot-dashed) curve, $(0,0.2)$ purple (dotted) curve and $(-0.1,0.1)$ brown (large double dot-dashed) curve. Bottom panel: $n_r=1.0,q=0.5,\alpha_r=-0.1,\beta_r=0.1$ for $(\alpha_\perp,\beta_\perp)=(0,0)$ blue (solid) curve, $(0.5,0)$ orange (dashed) curve, $(1.5,0)$ green (dot-dashed) curve, $(0.5,0.003)$ red (short double dot-dashed) curve, $(1.5,0.004)$ purple (dotted) curve and $(2.0,0.001)$ brown (large double dot-dashed) curve.}
    \label{fig3}
\end{figure*}

The parameter $y$, which is an observable related to the redshift of the object's surface (``the surface potential''), and can be used to measure the degree of compactness, is plotted in Fig. \ref{fig3} as a function of the anisotropy parameter $\theta$ \cite{6p} for different pairs of parameters as shown in the legend of the figure. From expression (\ref{M-Tolman}) we see that, in fact, we need the values of $y$ to obtain the behavior for the normalized Tolaman mass $\eta_T$ shown in Fig. \ref{fig2}. All curves show that the compactness of the object ($y$) decreases when the $\theta$ parameter grows. We observe that we have considerable changes in the magnitude of the compactness parameter of the relativistic sphere associated with changes in the surface potential as the set of parameters associated with the master polytrope varies and this may be an advantageous fact of our model since this potential is related to the redshift at the surface. Specifically, the potential decreases with the increasing of the radial polytrope index $n_r$. In contrast, $y$ increases, when the relativistic parameter $q_0$ grows as can be seen in the left panel of the first row in Fig. \ref{fig3}, where the doublet ($n_r$, $q_0$) were systematically considerate. So, if we stick to a specific polytropic model (fixed $n_r$) and increase $q_0$ the compactness will increase (as expected) and therefore the potential $y$, as we see for the red, purple, and brown curves. In the top right panel, the results are clearly visible showing the dependence of $y$ with the specific variations of the radial parameters ($\alpha_r$, $\beta_r$). In the second row of Fig. (\ref{fig3}), the same is done, fixing ($\alpha_r$, $\beta_r$) while changing separately the tangential parameters ($\alpha_\perp$, $\beta_\perp$) and in this way achieve a complete study of the properties (or possible advantages) of the master double polytrope. Different values for these parameters can produce noticeable effects that could be measured observationally and that can distinguish this model from the usual double polytrope (blue line) \cite{6p}.

\begin{figure*}
    \resizebox{1 \textwidth}{!}{
    \includegraphics{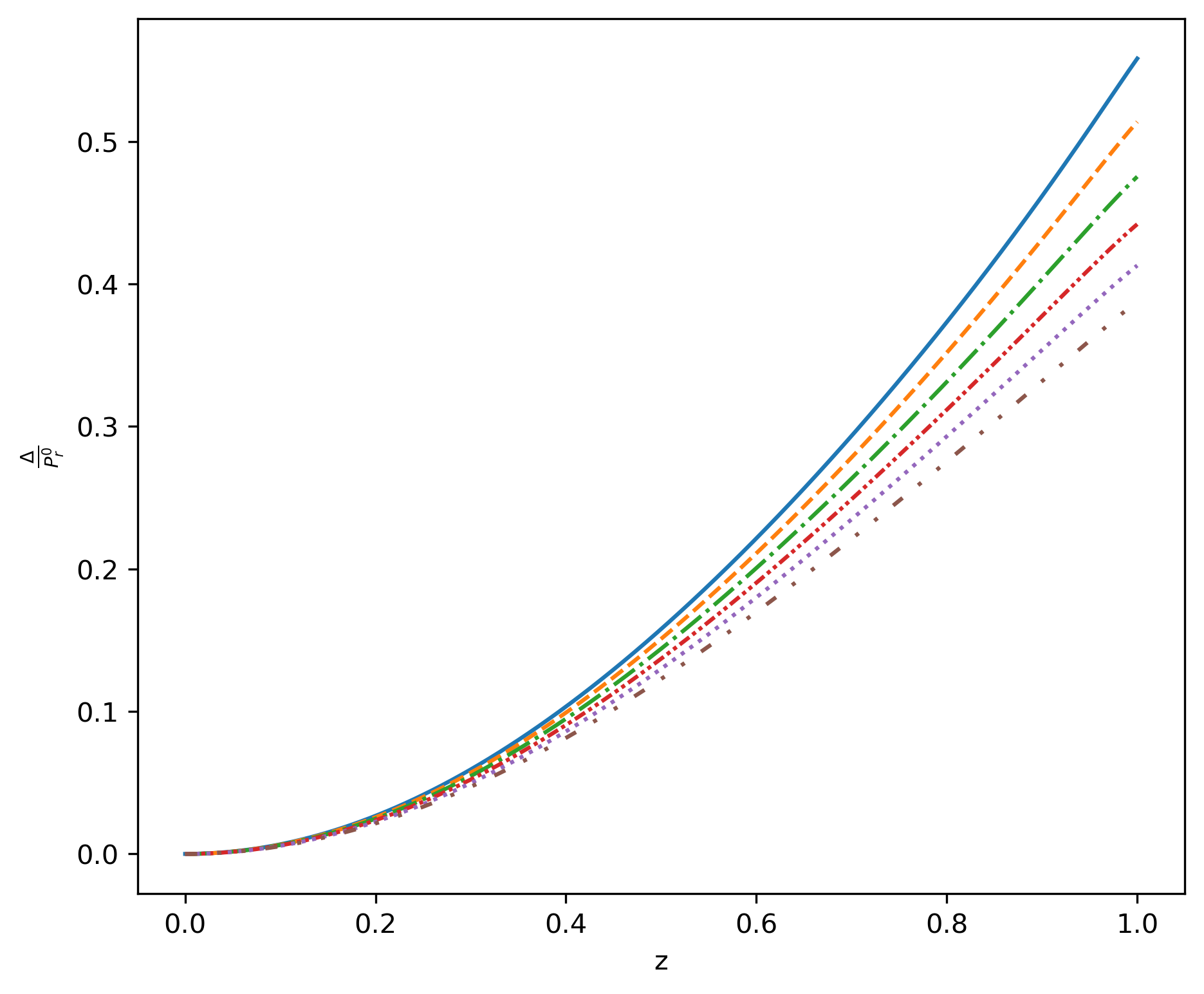} \ \includegraphics{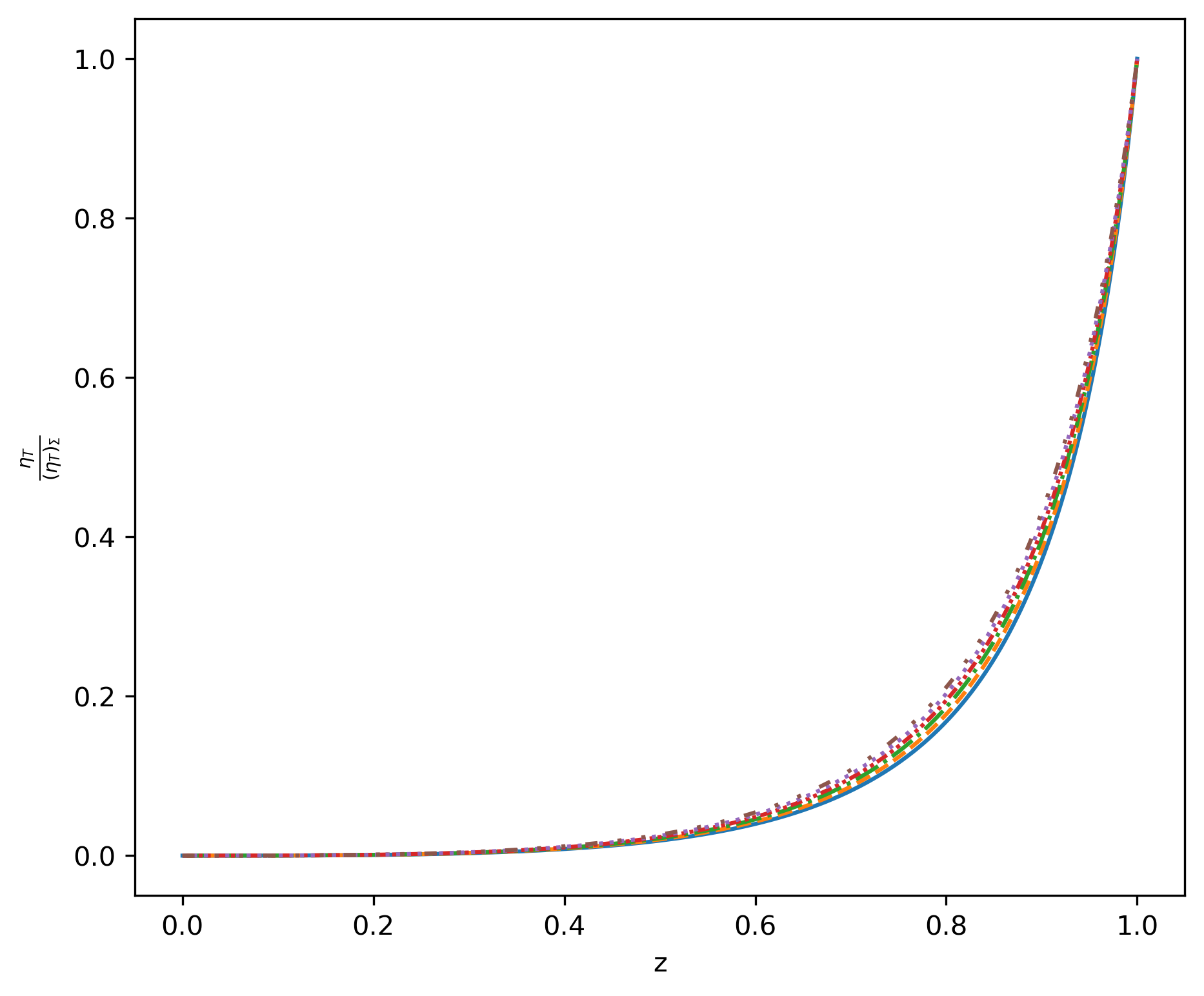}}
    \caption{$\Delta/P_r^0$ (left) and $\eta_T /(\eta_T)_\Sigma$ (right) as a function of $z$ for $n_r = 0.1$ $q_0 = 1.3$, $\alpha_r =-0.15$, $\beta_r=0.08$, $\alpha_\perp = 0.15$, $\beta_\perp=0$ and $\theta = 0.2$ blue (solid) curve, $\theta = 0.3$ orange (dashed) curve, $\theta = 0.4$ green (dot-dashed) curve, $\theta = 0.5$ red (short double dot-dashed) curve, $\theta = 0.6$ purple (dotted) curve and $\theta = 0.7$ brown (large double dot-dashed) curve.}
    \label{fig4}
\end{figure*}

From the expression (\ref{af}) we get that $\theta$ modulates the anisotropy of the system, for a specific polytrope configuration, and we plotted this dependency in the left panel of Fig. \ref{fig4}. As the parameter $\theta$ increases (for a specific polytrope given by the index $n_r$), the anisotropy decreases for the whole object although, near the center, the variation is less abrupt. Although this fact, certainly, is subject to the choice of the parameters ($\alpha_r$, $\beta_r$; $\alpha_\perp$, $\beta_\perp$), the behavior of the curves is qualitatively the same for a wide range of values of the parameters involved. In the right panel (of the same figure) we relate the normalized Tolman mass  $\eta_T /(\eta_T)_\Sigma$ with the $\theta$ parameter. There is no noticeable effect of the system to diminish the Tolman mass in the inner regions and to concentrate it in the outer ones as it does in the usual double polytrope. Virtually no effect is observed near the center or near the surface of the object.

Due to the fact that our model is based on a double polytrope that complies with a generalized master equation of state, in general, adequate behaviors for the functions involved are possible, but their reliance on ($\alpha_\perp$, $\beta_\perp$) is significantly more sensitive. Indeed, even for small values of $\beta_\perp$, there are cases in which we could not find any solution for the Lane-Emden system of equations. In the cases where we find a valid solution the behavior of the metric and thermodynamic functions becomes very similar to the one presented in Fig. \ref{fig1}. The effect of ($\alpha_\perp$, $\beta_\perp$)  will be analyzed in more detail in the next section.

\section{Acceptability conditions of the model}

\begin{figure*}
    \resizebox{1.0 \textwidth}{!}{
    \includegraphics{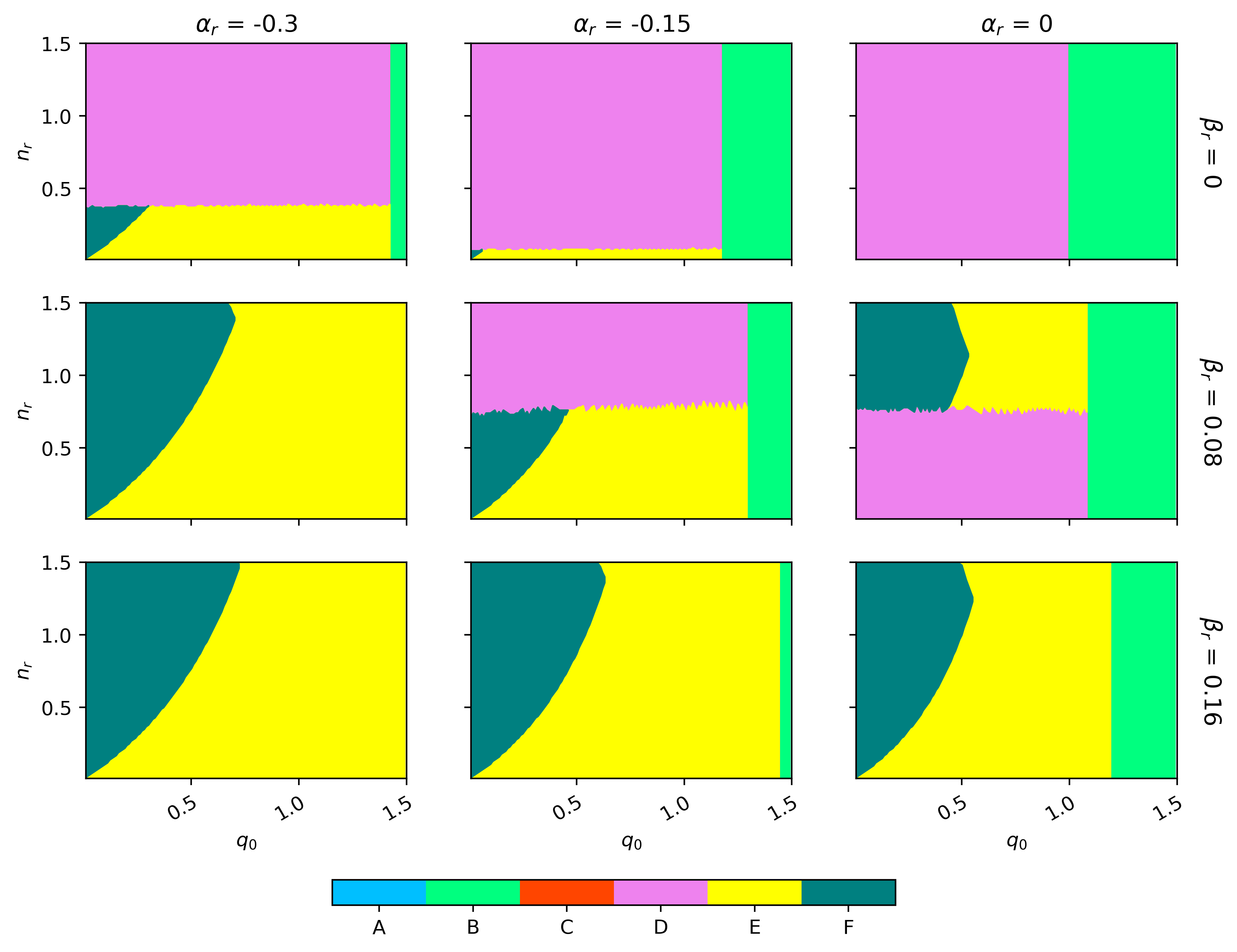}}
    \caption{Tuples ($n_r , q_0$) that satisfy the conditions 1-6 for $\alpha_\perp = \beta_\perp = 0$, $\theta = 0.9$. Each color in this graphic represents a set of conditions that are satisfied. These are: blue region (A) only condition 1, green region (B) conditions 1-2, red region (C) conditions 1-3, violet region (D) conditions 1-4, yellow region (E) conditions 1-5 and dark green region (F) conditions 1-6. }
    \label{fig5}
\end{figure*}

\begin{figure*}
    \resizebox{1.0 \textwidth}{!}{
    \includegraphics{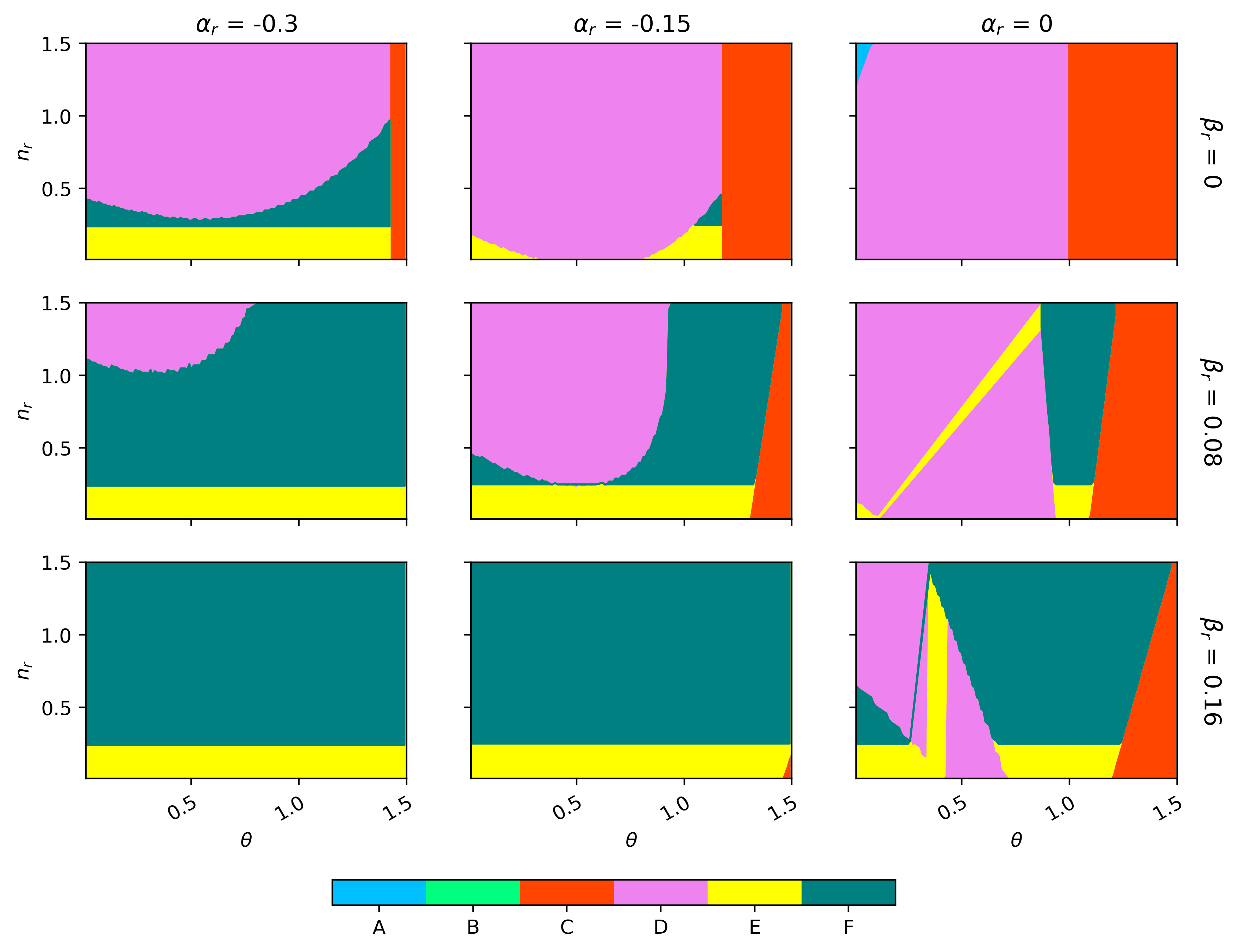}}
    \caption{Tuples ($n_r ,\theta$) that satisfy the conditions 1-6 for $\alpha_\perp = \beta_\perp = 0$, $q_0 = 0.2$. Each color in this graphic represents a set of conditions that are satisfied. These are: blue region (A) only condition 1, green region (B) conditions 1-2, red region (C) conditions 1-3, violet region (D) conditions 1-4, yellow region (E) conditions 1-5 and dark green region (F) conditions 1-6. }
    \label{fig6}
\end{figure*}

\begin{figure*}
    \resizebox{1.0 \textwidth}{!}{
    \includegraphics{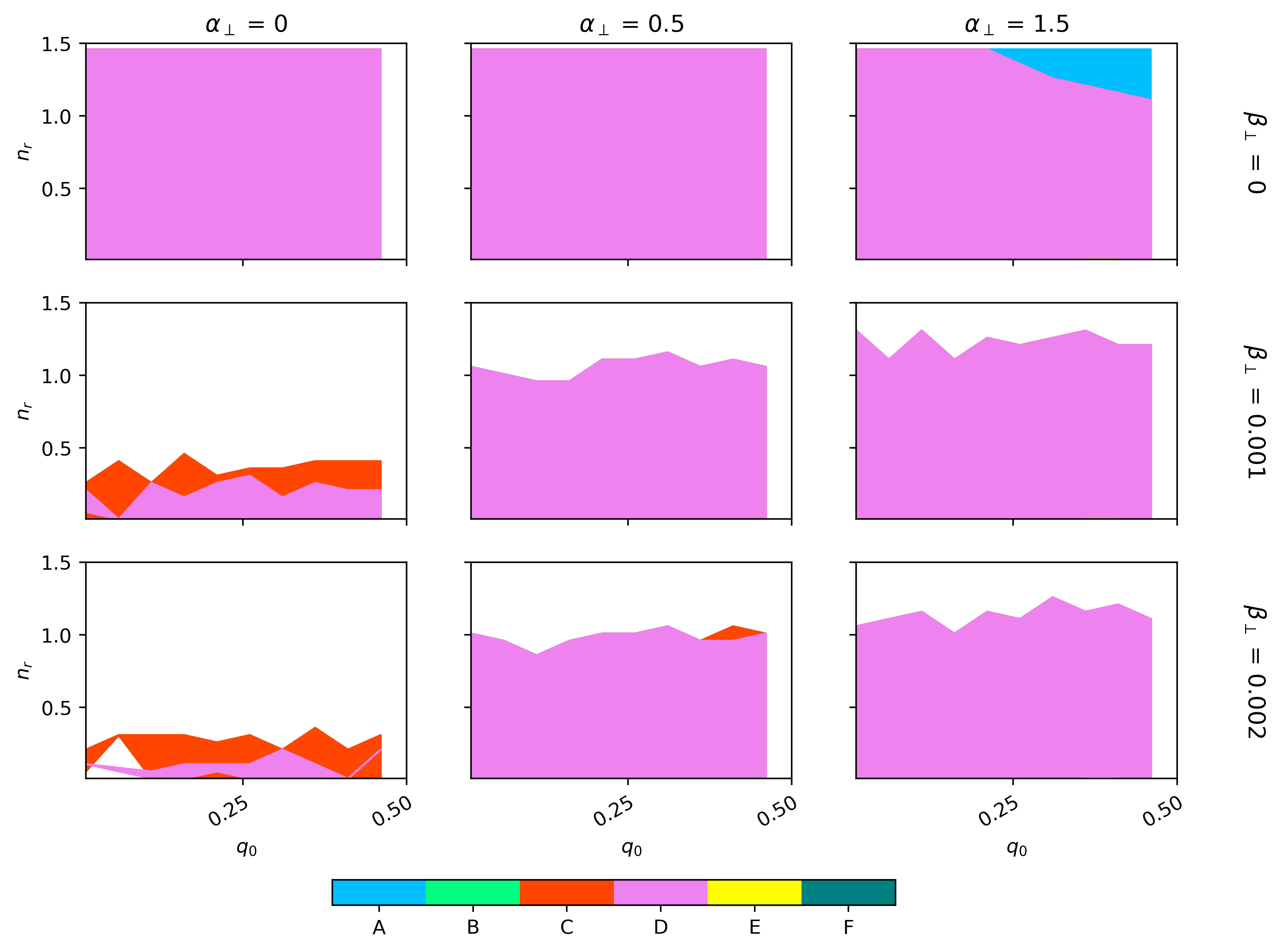}}
    \caption{Tuples ($n_r , q_0$) that satisfy the conditions 1-6 for $\alpha_r = \beta_r = 0$, $\theta = 0.9$. Each color in this graphic represents a set of conditions that are satisfied. These are: blue region (A) only condition 1, green region (B) conditions 1-2, red region (C) conditions 1-3, violet region (D) conditions 1-4, yellow region (E) conditions 1-5 and dark green region (F) conditions 1-6. }
    \label{fig7}
\end{figure*}

\begin{figure*}
    \resizebox{1.0 \textwidth}{!}{
    \includegraphics{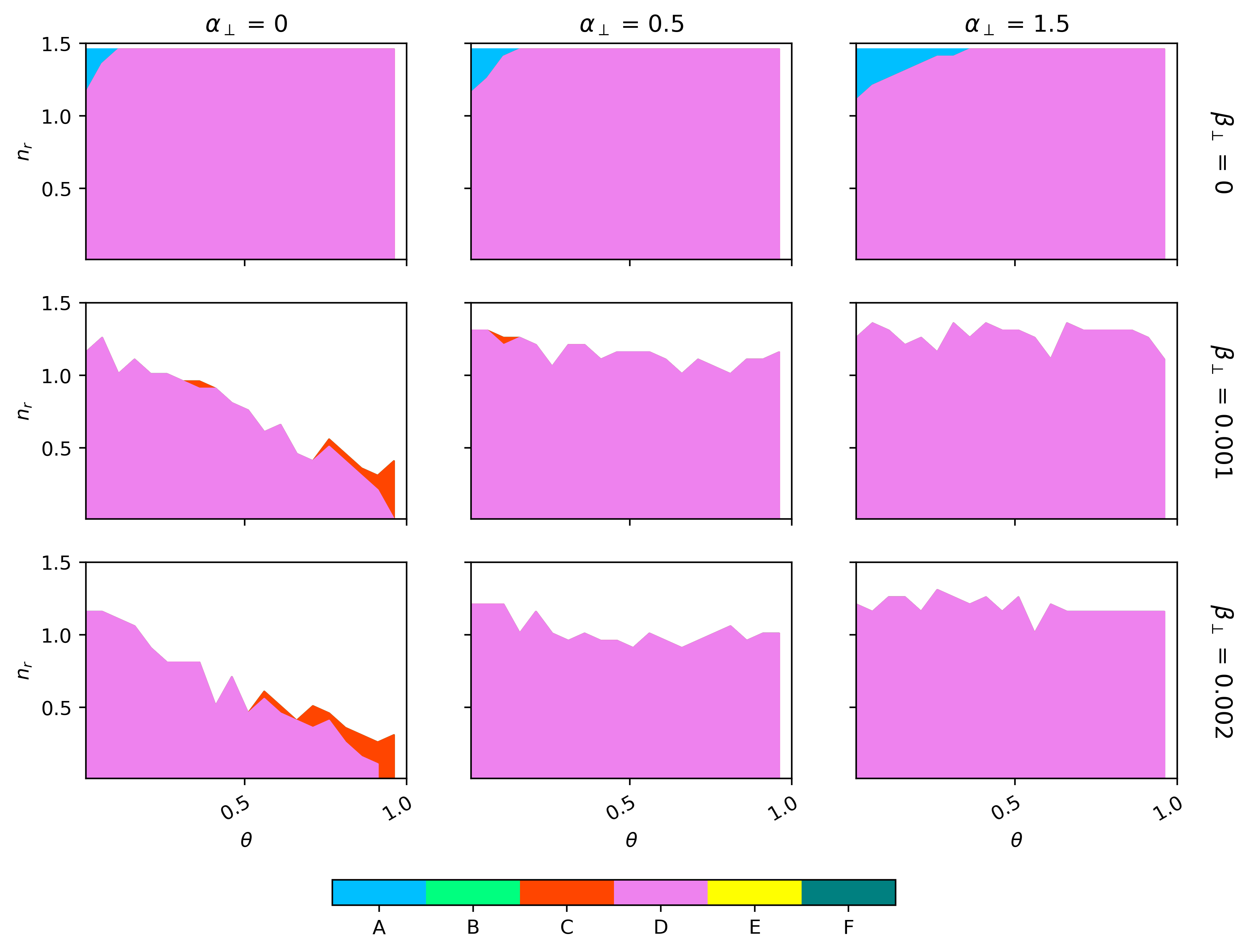}}
    \caption{Tuples ($n_r , \theta$) that satisfy the conditions 1-6 for $\alpha_r = \beta_r = 0$, $q_0 = 0.2$. Each color in this graphic represents a set of conditions that are satisfied. These are: blue region (A) only condition 1, green region (B) conditions 1-2, red region (C) conditions 1-3, violet region (D) conditions 1-4, yellow region (E) conditions 1-5 and dark green region (F) conditions 1-6. }
    \label{fig8}
\end{figure*}

In order to explore the effects of the new variables $\alpha_r,\beta_r,\alpha_\perp,\beta_\perp$, we check the set of tuples $(n_r,q_0)$ and $(n_r,\theta)$ that satisfy the following conditions:

\begin{enumerate}
    \item The thermodynamic variables $\rho,P_r$ and $P_\perp$ are positive and finite at the center of the configuration with $P_r = P_\perp$.
    \item $\rho,P_r$ and $P_\perp$ are monotonously decreasing functions, having the maximum at the center of the configuration: $\rho'_{c} = P'_{r c} = P'_{\perp c} = 0$.
    \item The strong energy conditions $\rho + P_r + 2P_\perp \geq 0$, for imperfect fluids \cite{Ivanov}. 
    \item The anisotropy function $\Delta = P_\perp - P_r$ is positive.
    \item The anisotropy function $\Delta$ is a monotonously increasing function. 
    \item The sounds velocities satisfy the causality conditions $0<v^s_r \leq 1$ and $0<v^s_\perp \leq 1$ \cite{cn4, Delgaty}.
\end{enumerate}
In figures (\ref{fig5})-(\ref{fig8}) we show our results regarding the points listed above. Figures (\ref{fig5})-(\ref{fig6}) represent different values of $\alpha_r$ and $\beta_r$ with $\alpha_\perp = \beta_\perp = 0$ while figures (\ref{fig7})-(\ref{fig8}), represent different values of $\alpha_\perp$ and $\beta_\perp$ with $\alpha_r = \beta_r = 0$. From these figures we can summarize the following points:
\begin{itemize}
    \item We found that negative values of $\alpha_r$ slightly increase the number of matter distributions satisfying all the conditions 1-6. This is clear from figures (\ref{fig5}) and (\ref{fig6}). Especially in the top panels. 
    \item We see that positive values of $\beta_r$ lead to a huge improvement in the number of matter distributions satisfying all the conditions 1-6. This is clear also from figures (\ref{fig5}) and (\ref{fig6}). 
    \item Analyzing the figures (\ref{fig7}) and (\ref{fig8}) we observe that positives values of $\alpha_r$ with $\beta_r=0$ leads, in general, to a worse behavior. However, we found an improvement in this when both $\alpha_r > 0$ and $\beta_r > 0$.
    \item In general, negative values of $\beta_r$ lead to worse behaviors for compact object fluid distribution.
    \item Smallest values of $q_0$ lead to better behavior, in particular for the conditions 5-6. 
    \item In general, the increase of the $\theta$ parameter (used to ``control'' the anisotropy) up to a certain value helps to fulfill the conditions, having beyond that value the opposite effect. A certain amount of anisotropy can contribute to stability and better behavior of stars and this seems to be crucial in the proper behavior of relativistic compact objects \cite{cn4}.
    \item For $n_r$ we found that smaller values lead to better behaviors to satisfy the matter conditions for the fluid distributions, at least for conditions $1-5$.  However, in some cases, there is a peculiar behavior that leads to the fact that the condition $6$ is broken for small values of $n_r$ (see figures (\ref{fig5}) and (\ref{fig6})).
    \item For $\alpha_r = 0 $ and $\beta_r=0$ and the values of $\alpha_\perp, \beta_\perp$ considered in this work, we could not find any case in which conditions $5-6$ were satisfied, as seen in figures (\ref{fig7}) and (\ref{fig8}). This scenario improves when both  $\alpha_r\not=0$ and $\beta_r\not=0$, in particular, is accentuated for the last. However, as can be checked by the reader, the behavior only change slightly with respect to the shown in Figs. (\ref{fig7}) and (\ref{fig8}) so we do not present these results here.

    \item Positive values of $\alpha_\perp$ may improve the behavior of the solutions, but it seems that only affect the conditions 1-4 and not 5-6. 
    
    \item In general, we obtain that  $\beta_\perp\not= 0$ decreases the number of cases in which conditions $1-6$ are satisfied. For small values of $\beta_\perp$ the behavior may improve with the inclusion of positive values of $\alpha_\perp,\beta_r$ and negative values of $\alpha_r$. However, if we continue increasing $\beta_\perp$, the negative effects of this constant will be dominant and this trend does not improve appreciably with the inclusion of the other constants.
    
\end{itemize}

\section{Cracking of the master double polytrope}

In this section, we will use the perturbation scheme presented in section \ref{section3} to find an expression for the total radial force after perturbations of the energy density and local pressure anisotropy are performed. In order to achieve such a goal, we introduce Eqs. (\ref{Pr}) and (\ref{af}) in (\ref{Rfinal}) obtaining
\begin{eqnarray}
\hat{R} & = & q_0[\omega^{n_r}(\omega+\alpha_r)-\beta_r]\Bigg[\frac{b(x)x^2}{c(x)} + \frac{d(x)}{xc(x)} \times \nonumber\\ && \qquad\qquad\qquad\qquad \Bigg(1 + \frac{n_r}{q_0(\omega(n_r+1)+n_r \alpha_r)}\Bigg) \Bigg]\nonumber\\ &+&\frac{b(x)f(x)\tilde{F}_1}{c(x)^2} + \frac{1}{(1+n_r)} \omega^{n_r} \left(n_r+1+\frac{n_r\alpha_r}{\omega}\right)\frac{d\omega}{dx}  \nonumber \\
&-& \frac{2}{(1+n_r)x \Gamma} [a(\omega^{n_r}(\omega^\theta +\alpha_\perp)-\beta_\perp)\nonumber\\ && 
\qquad\qquad\qquad\qquad\qquad\; -\omega^{n_r}(\omega+\alpha_r)+\beta_r],
\end{eqnarray}
where
\begin{eqnarray*}
\tilde{F}_1 &=& \int_0^x \Bar{x}^2 \left(\frac{\omega}{n_r}(n_r +1) + \alpha_r\right)^{-1}(\omega^{n_r}(\omega +\alpha_r)-\beta_r)d\Bar{x}, \\
b(x) &\equiv& \omega^{n_r}+q_0(\omega^{n_r}[\omega + \alpha_r]-\beta_r), \\
c(x) &\equiv& x-2(1+n_r)q_0\eta, \\
d(x) &\equiv& \eta + x^3 q_0(\omega^{n_r}[\omega + \alpha_r]-\beta_r), \\
f(x) &\equiv& 1+2(1+n_r)q_0^2 x^2(\omega^{n_r}[\omega +\alpha_r]-\beta_r)
\end{eqnarray*}
and 
\begin{eqnarray}
\hat{R} \equiv \frac{\tilde{R}}{4\pi \rho^2_c A \delta \phi}.
\end{eqnarray}

\begin{figure*}
    \centering
    \begin{subfigure}[b]{0.49 \textwidth}
    \includegraphics[width=\textwidth]{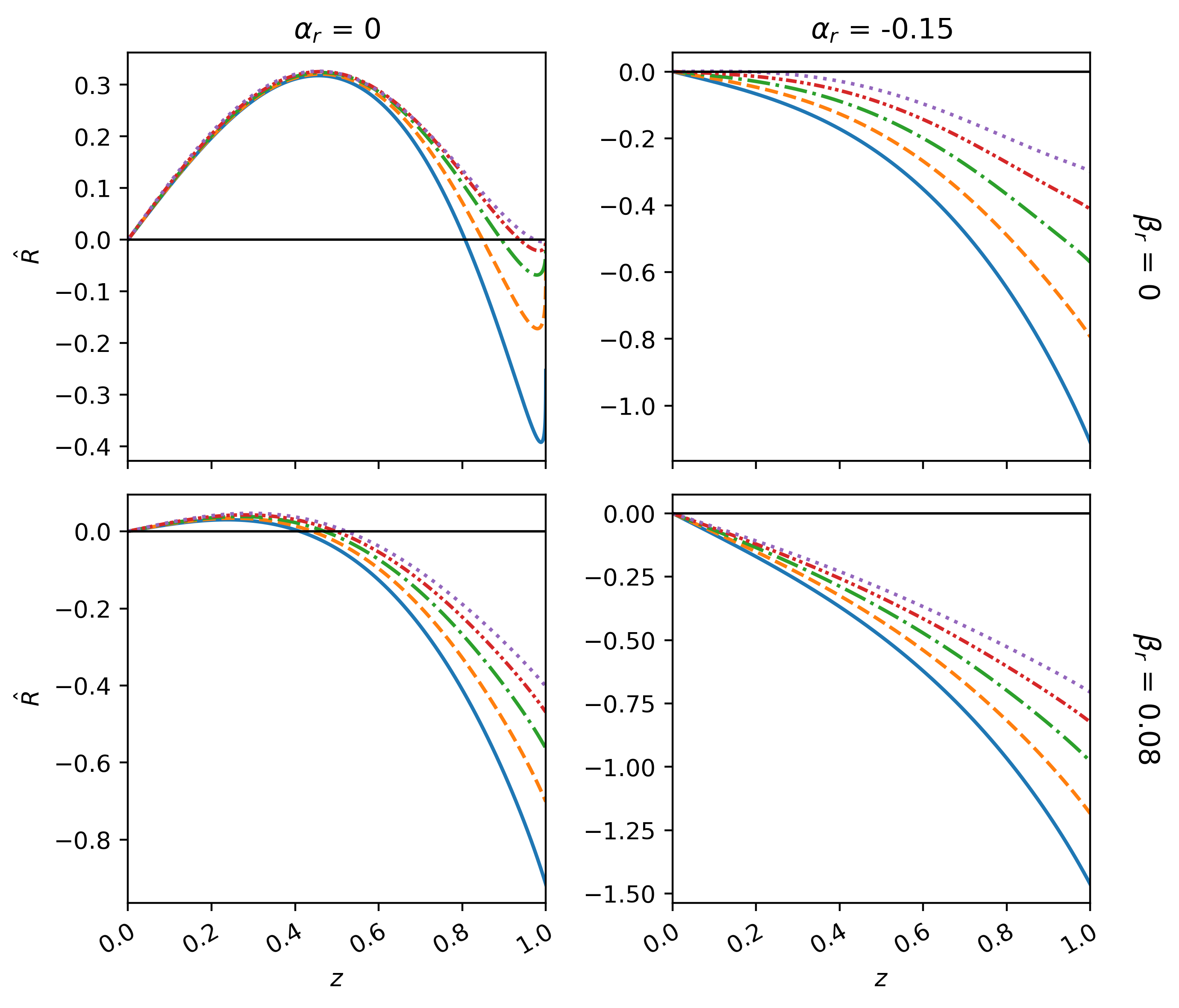}
    \caption{}
    \end{subfigure}
     \hfill
     \begin{subfigure}[b]{0.49 \textwidth}
      \includegraphics[width=\textwidth]{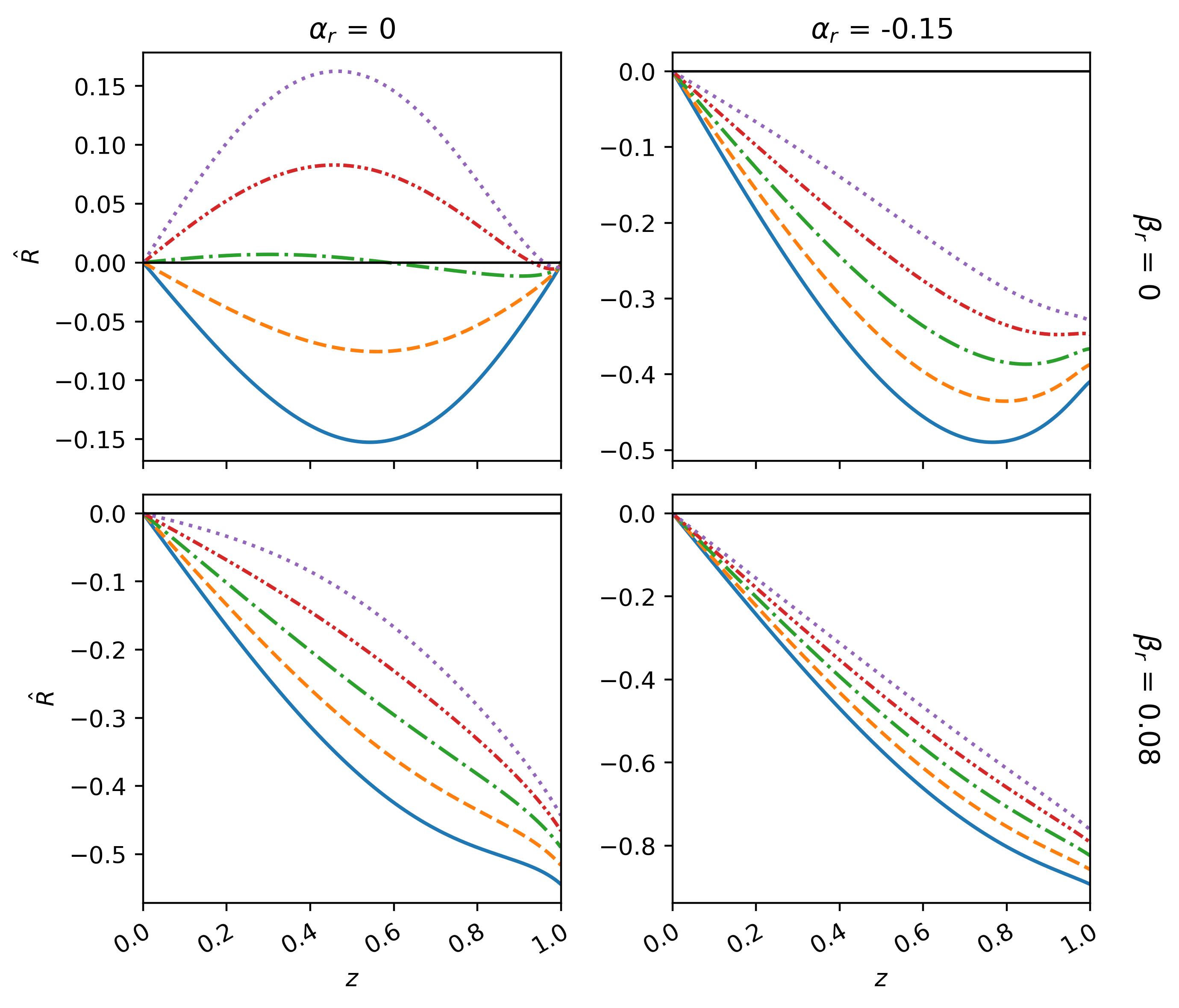}
     \caption{}
     \end{subfigure}
     \caption{(a) $\hat{R}$ vs $z$ for $q_0=0.2$, $\theta = 1.2$, $\Gamma = 0.15$, $\alpha_\perp=\beta_\perp=0$ and $n_r=0.1$ blue (solid) curve, $n_r=0.2$ orange (dashed) curve, $n_r=0.3$ green (dot-dashed) curve, $n_r=0.4$ red (short double dot-dashed) curve, $n_r=0.5$ purple (dotted) curve. (b) $\hat{R}$ vs $z$ for $q_0=0.2$, $n_r= 0.5$, $\Gamma = 0.15$, $\alpha_\perp=\beta_\perp=0$ and $\theta=0.90$ blue (solid) curve, $\theta=0.95$ orange (dashed) curve, $\theta=1.00$ green (dot-dashed) curve, $\theta=1.05$ red (short double dot-dashed) curve, $\theta=1.10$ purple (dotted) curve.}
    \label{fig11}
\end{figure*}

\begin{figure*}
    \centering
    \begin{subfigure}[b]{0.49 \textwidth}
    \includegraphics[width=\textwidth]{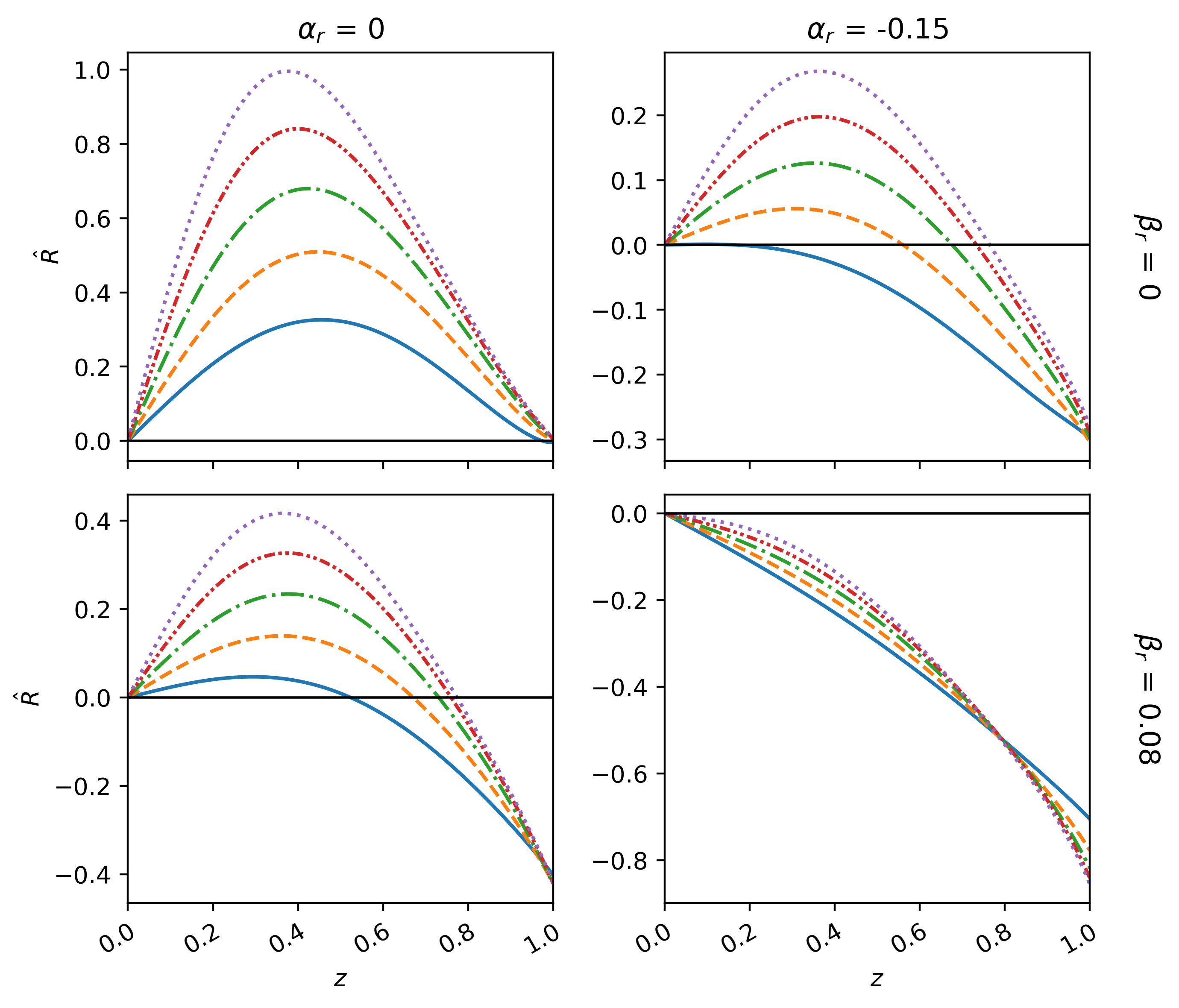}
    \caption{}

    \end{subfigure}
     \hfill
     \begin{subfigure}[b]{0.49 \textwidth}
      \includegraphics[width=\textwidth]{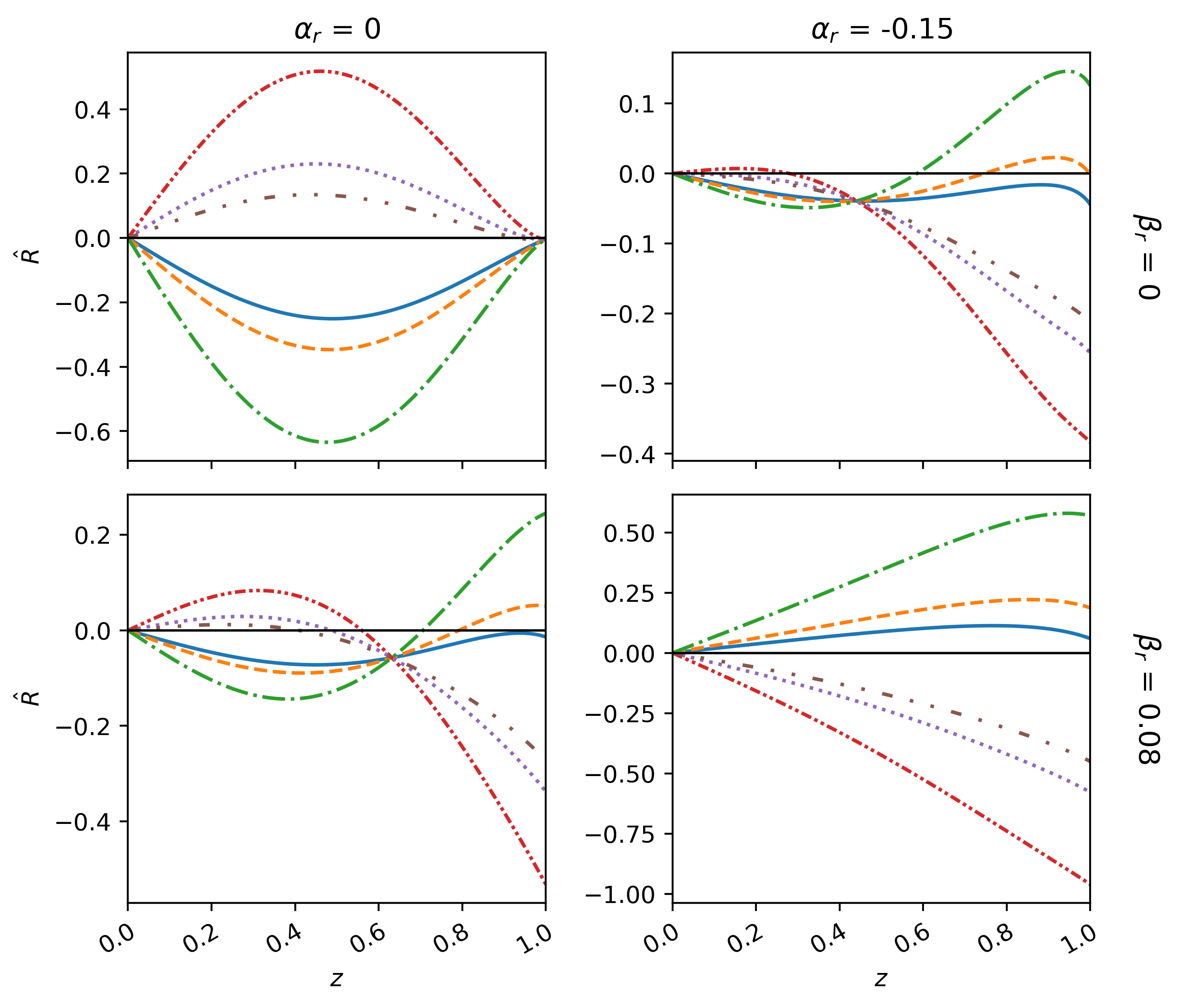}
     \caption{}
     \end{subfigure}
     \caption{(a) $\hat{R}$ vs $z$ for $n_r=0.5$, $\theta = 1.2$, $\Gamma = 0.15$, $\alpha_\perp=\beta_\perp=0$ and $q_0=0.2$ blue (solid) curve, $q_0=0.4$ orange (dashed) curve, $q_0=0.6$ green (dot-dashed) curve, $q_0=0.8$ red (short double dot-dashed) curve, $q_0=1.0$ purple (dotted) curve. (b) $\hat{R}$ vs $z$ for $q_0=0.2$, $n_r= 0.2$, $\theta = 1.2$, $\alpha_\perp=\beta_\perp=0$ and $\Gamma=-0.3$ blue (solid) curve, $\Gamma=-0.2$ orange (dashed) curve, $\Gamma=-0.1$ green (dot-dashed) curve, $\Gamma=0.1$ red (short double dot-dashed) curve, $\Gamma=0.2$ purple (dotted) curve and $\Gamma=0.3$ brown (large double dot-dashed) curve.}
    \label{fig12}
\end{figure*}

\begin{figure}
    \resizebox{0.5 \textwidth}{!}{
    \includegraphics{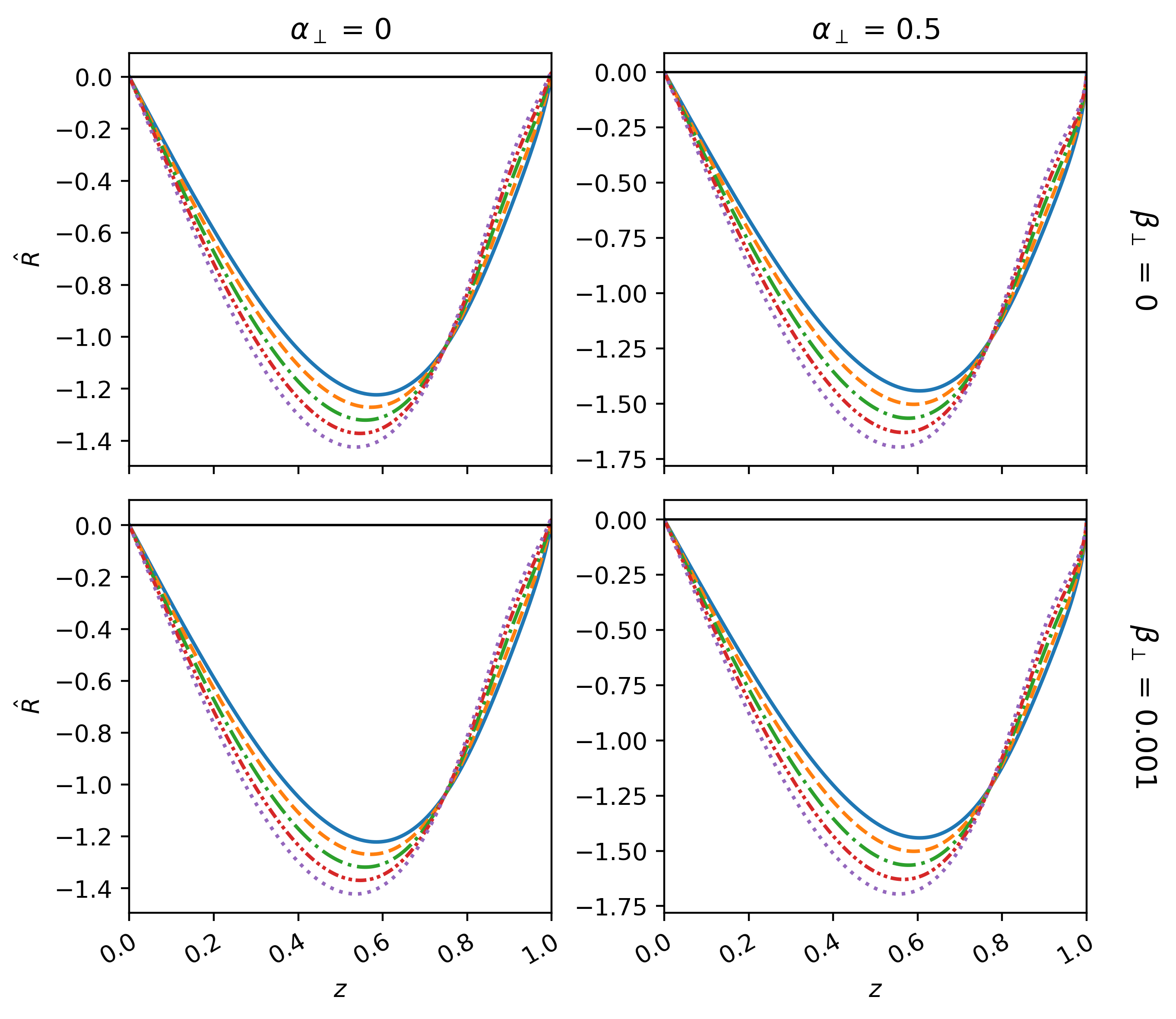}}
    \caption{$\hat{R}$ vs $z$ for $q_0=0.2$, $\theta = 1.2$, $\Gamma = 0.15$, $\alpha_r=\beta_r=0$ and $n=0.1$ blue (solid) curve, $n_r=0.2$ orange (dashed) curve, $n_r=0.3$ green (dot-dashed) curve, $n_r=0.4$ red ( short double dot-dashed) curve, $n_r=0.5$ purple (dotted) curve.}
    \label{fig13}
\end{figure}

In figures \ref{fig11},\ref{fig12}, and \ref{fig13}, we show $\hat{R}$ as a function of $z$ for the different values of the parameters involved, specified in the legend of the figures.

\begin{itemize}
    \item The left panel of the first row in Figs. \ref{fig11} (a) and \ref{fig12} (a)
corresponds to the case of the usual double polytrope \cite{6p} and coincides, as it should, with the study of cracking presented in \cite{efl}. 
    \item From figures \ref{fig11} and \ref{fig12} it is clear that $\alpha_r$ and $\beta_r$ have a great impact on the total radial force, and therefore in the apparition of cracking, after the system is perturbed. They are relevant parameters associated with this fact. The results indicate that negative values for $\alpha_r$ in combination with slightly positive values for $\beta_r$ have some influence in avoiding (see Fig. \ref{fig11}) or promoting (see Fig. \ref{fig12}) cracking.
    \item In figures \ref{fig11} and \ref{fig12} we can see that bigger values of $\beta_r$ along with the most negative values of $\alpha_r$ have the effect of gradually changing the radial force direction compared to the case were $\alpha_r=\beta_r=0$. This is very clear from Fig. \ref{fig12} (b). Thus, in the case where cracking is present for $\alpha_r=\beta_r=0$, the change of sing will take place in the deeper regions of the fluid for bigger positive values of $\beta_r$ and more negative values of $\alpha_r$. In the case of overturning the contrary holds. 
    \item In this sense, we can have cases where there is no cracking (overturning) when $\alpha_r=\beta_r=0$ but it may appear when $\alpha_r=\beta_r\not=0$. The opposite is also possible. 
    \item In figure \ref{fig13} we isolate the role played by the tangential parameters $\alpha_\perp$ and $\beta_\perp$ related to our master relativistic double polytrope model. It can be observed that they do not have a big impact on the total radial force, at least for the values used in the previous section. We found the same results for different values of $q_0$, $\theta$, and $\Gamma$. These aspects, described in the last two items, are considered to be interesting peculiarities that our model presents and deserve special attention since they could be related to ``stability'' facts against cracking that relativistic fluids present. 
    \item We found that bigger values for the tangential pressure parameters $\alpha_\perp$ and $\beta_\perp$ may affect the radial force. However, from the results of the previous section, this will lead to ill-behaved matter distributions. Thus small variations of $\alpha_\perp$ have a great impact on the behavior of the thermodynamics variables but not in the radial force just after the system departs from hydrostatic equilibrium.
\end{itemize}

\section{Discussion}

This work may be interpreted as a natural generalization of the approach described in \cite{6p}, in the relativistic regime, building an anisotropic model where it is assumed that both pressures satisfy the polytrope master equation, which has some versatility and could be applied to other types of scenarios. Modeling compact objects with anisotropic polytropes started in 2013 [78] and generated many exciting candidates. The fact of using a polytropic equation of state to describe compact objects has been carried out for several years producing some interesting candidates. Also, due to the great relevance that anisotropic internal solutions have acquired in recent years in the structure of self–gravitating objects, and by the fact that polytropes represent fluid systems with a wide range of applications in astrophysics as Fermi fluids, super-Chandrasekhar white dwarfs, we have described hereby a general framework for modeling of general relativistic polytropes in the presence of anisotropic pressure, when both pressures satisfy a master polytropic equation of state. Furthermore, as a theoretical motivation for our work, we have the fact that polytropes have been very successful in describing astrophysical objects such as white dwarfs and neutron stars (in both, Newtonian and relativistic realms) and constitute a versatile way of obtaining the Chandrasekhar limit value (and even the upper limit for the mass of stars composed of degenerate neutron matter). It is worth emphasizing that some of the physical phenomena present in such configurations (e.g. very strong magnetic fields) could break the spherical symmetry, implying thereby that our approach should be taken, in this case, as an approximation.

The reason to adopt such an assumption is provided by the simple fact that for small anisotropies it is always a good approximation and also because of the wealth provided by using both equations of state of a master polytrope for each pressure. This type of equation of state (the master polytrope) has been very versatile to include several particular cases found in the literature (see for example the references \cite{8lp, Nasheeha}). The variation of the variables involved generates a parameter space representing a wide range of possible astrophysical candidates \cite{LN2}. Our case associated with a double master polytrope presents a greater number of parameters ($n_r$, $q_0$, $\theta$, $\alpha_r$, $\beta_r$, $\alpha_\perp$, $\beta_\perp$, etc.) with interesting behaviors when studying possible realistic configurations. Note that, although the double master polytrope is not an equation of state that comes from some known thermodynamic process, there are a couple of reasons to consider this equation of states that motivate their use in this work. First, note that the master equation of state was used in Ref. (\cite{8lp})  with the aim to avoid some pathologies that, exist in all polytropic non-Pascalian fluids. Namely, it is known that those models having $\alpha=\beta=0$ with $1<\gamma<2$ present a singularity in the tangential sound velocity at the boundary of the matter distribution. Second, by considering the double polytropic equation we ensure that the system is closed enough in the sense that as both the tangential and the radial pressure fulfill an equation of state, we do not need to provide any extra (and sometimes arbitrary) information.

Also, we have investigated the conditions under which, general relativistic double master polytropes, exhibit cracking (and/or overturning), when submitted
to fluctuations of energy density and anisotropy. To achieve this, we used the general and systematic method proposed in \cite{efl} to study the departure from equilibrium for any internal, anisotropic, and spherically symmetric solution of Einstein field equations. Thus, we have shown that cracking occurs for a wide range of the parameters and the main conclusions are basically the determining role played by $\alpha_r$ and $\beta_r$ to create or avoid cracking, and the very small impact of the role of the same variables associated with the tangential pressure $P_\perp$. As previously stated in the preceding sections, the impact of the new parameters on the occurrence or avoidance of cracking will play a role in the structure and evolution of the systems presented here on a time scale smaller than the hydrostatic time scale. In order to obtain more information about the system over a longer time scale, it will be necessary to integrate the full dynamical field equations, which is outside the scope of our analysis.

Although the purpose of this work is not to model any particular astrophysical object, we would like to call attention to the potential application or connection of the approach presented here to interpret and explain some aspects concerning super-Chandrasekhar white dwarfs. These stars may attain masses of the order of $2.8M_\odot$ and are modeled resorting to a polytropic equation of state (see \cite{Kalita} and references therein). Now, given that our model has a great variety of free parameters, we are tempted to think that our results could possibly fit the observational data of these unusual configurations. We must stress that it is not yet fully understood why these distributions violate the upper bound mass for the white dwarfs. Different theories have been proposed to explain this phenomenon, however, these massive stars possess very little luminosity, and hence cannot be detected directly by any observations. For each of them, it is evident that general relativistic effects as well as the inclusion of pressure anisotropy, are unavoidable. Nevertheless, care must be exercised with the fact that some of the physical phenomena present in such configurations (e.g. the presence of magnetic field and rotation) could break the spherical symmetry, implying thereby that our approach should be taken as an approximation.

\section{Acknowledgements}

P.L wants to say thanks for the financial support received by the Projects ANT1956 and ANT2255 of the Universidad de Antofagasta. P.L is grateful Semillero de Investigaci\'on SEM 18-02 from Universidad de Antofagasta.

\end{document}